\documentclass[a4paper,11pt]{article}
 \pdfoutput=1
 \usepackage{jheppub} 
 \usepackage{graphicx}
 \usepackage{amssymb}
\usepackage{dcolumn}% Align table columns on decimal point
\usepackage{bm}% bold math 
\usepackage{color}
\usepackage[table]{xcolor}
\usepackage{multirow}
\usepackage{soul}

\allowdisplaybreaks

 \newcommand{\lsim}{{\;\raise0.3ex\hbox{$<$\kern-0.75em\raise-1.1ex\hbox{$\sim$}}\;}}
\newcommand{\gsim}{{\;\raise0.3ex\hbox{$>$\kern-0.75em\raise-1.1ex\hbox{$\sim$}}\;}}
\newcommand{\beq}{\begin{equation}}
\newcommand{\eeq}{\end{equation}}
\newcommand{\bea}{\begin{eqnarray}}
\newcommand{\eea}{\end{eqnarray}}
\mathchardef\minus="002D

\def\gev{\textrm{ GeV}}
\def\tev{\textrm{ TeV}}

\def\beq{\begin{equation}}
\def\eeq{\end{equation}}
\def\bea{\begin{eqnarray}}
\def\eea{\end{eqnarray}}
\def\bit{\begin{itemize}}
\def\eit{\end{itemize}}

\def\baa{\begin{array}}
\def\eaa{\end{array}}

\usepackage{cancel}
\def\misse{\cancel{p}_{T}}
\def\missev{\cancel{\vec{p}}_{T}}

\title{\boldmath Mass Measurement Using Energy Spectra in Three-body Decays}

\author[a]{Kaustubh Agashe}  
\author[a,b]{Roberto Franceschini} 
\author[a,c]{Doojin Kim} 
\author[a]{Kyle Wardlow}

\affiliation[a]{Maryland Center for Fundamental Physics, Department of Physics, University of Maryland, College Park, MD 20742, USA}
\affiliation[b]{CERN Physics Department, Theory Division, CH-1211 Geneva 23, Switzerland.}
\affiliation[c]{Department of Physics, University of Florida, Gainesville, FL 32611, USA}

\preprint{UMD-PP-015-004, CERN-PH-TH-2015-033}
\emailAdd{kagashe@umd.edu}
\emailAdd{roberto.franceschini@cern.ch}
\emailAdd{immworry@ufl.edu}
\emailAdd{kwardlow@umd.edu}

\abstract{
In previous works we have demonstrated how the energy distribution of massless decay products in two body decays can be used to measure the mass of decaying particles. In this work we show how such results can be generalized to the case of multi-body decays.  The key ideas that allow us to deal with multi-body final states are an extension of our previous results to the case of massive decay products and the factorization of the multi-body phase space.
The mass measurement strategy that we propose is distinct from alternative methods because it does not require an accurate reconstruction of the entire event, as it does not involve, for instance, the missing transverse momentum, but rather requires measuring only the visible decay products of the decay of interest.  
To demonstrate the general strategy, we study a supersymmetric model wherein pair-produced gluinos each decay to a stable neutralino and a bottom quark-antiquark pair via an {\em off}-shell bottom squark. The combinatorial background stemming from the indistinguishable visible final states on both decay sides can be treated by an ``event mixing'' technique, the performance of which is discussed in detail.  Taking into account dominant backgrounds, we are able to show that the mass of the gluino {\em and}, in favorable cases, that of the neutralino can be determined by this mass measurement strategy.}

\begin{document} 
\maketitle
\flushbottom

%%%%%%%%%%
%%%%%%%%%%    Main Text
%%%%%%%%%%

\section{Introduction and general strategy of the mass measurement \label{sec:generalstrategies}}

%usual

A stable weakly interacting massive particle (WIMP), with a  weak-scale mass, is a  well-motivated
candidate for the dark matter (DM) of the universe. The reason for this is that it naturally results in the observed relic density after thermal freeze-out of the
early universe (for review, see for example, Ref.~\cite{Bertone:2004pz}).
In fact, 
this WIMP DM can also have weak-scale interactions with various Standard Model (SM) particles.\footnote{This is especially true if the WIMP DM 
arises as part of an extension to the SM that has been invoked to solve the Planck-weak hierarchy problem.} 
This feature leads to the exciting possibility that DM arising in this manner could be detected {\em non}-gravitationally\footnote{cf.
thus far, the only evidence for DM is via its gravitational effects.}.
In these cases, a symmetry under which the dark matter is charged must be present in order to stabilize this DM from decaying to purely SM final states; the SM particles remain uncharged, thus preventing decay.

While there are many possible non-gravitational probes of such DM,
%\footnote{\yellow{including detection of {\em ambient} DM}} 
we focus here on production of DM at hadron colliders, specifically the Large Hadron Collider (LHC).
Specifically, many models incorporating DM of the aforementioned type contain particles that are not only heavier than the DM and charged under the DM stabilization symmetry, but also that interact via SM gauge bosons.
If these heavier particles (dubbed ``parent" particles) do interact via say, QCD, they could be copiously produced at hadron colliders, which would then be followed by their subsequent decay into the concomitant DM and SM particles. By design, the DM particle leaves {\it no} visible trace in the particle detector, thus its presence in an event is typically inferred from the missing transverse momentum ($\misse$), which can be interpreted as a loss of specificity in the kinematic information of the event.

The primary goal of this paper is to devise a strategy for the simultaneous measurement of the masses of the parent {\em and} the DM particles in the associated processes despite this loss of information. This strategy for the mass measurement also has further applications beyond the study of DM particles; it can be applied to any case where a new particle decays to a semi-invisible final state. In fact, the invisible particle neither has to be a DM candidate, nor has to be absolutely stable -- only insofar as its time-of-flight out of the detector is concerned. However, for notational simplicity we shall still refer to it as ``DM".

A full reconstruction of such a decay chain is typically not possible, given that it contains an invisible particle. On top of this, due to the DM stabilization symmetry the parent particles are typically {\em pair}-produced, implying that each event comes with two invisible particles. The presence of two invisible particles involves an even greater loss of kinematic information from each event, and poses a sizable challenge in the associated mass measurement.
Methods using the $M_{ T2 }$ variable and its variations~\cite{Lester:1999et, Barr:2003fj, Cho:2007qv, Cho:2007dh, Barr:2008qy, Barr:2011ao} have been proposed as a solution to overcome this challenge. This class of variables is well-known for its usefulness both in measuring the masses of particles (e.g. Refs.~\cite{ATLAS-CONF-2012-082} and  \cite{Chatrchyan:2013boa}) and in isolating new physics signals from their backgrounds (e.g. Ref.~\cite{CMS-PAS-SUS-13-019}).
Despite their utility, these variables have a possible drawback when aiming at a precise mass measurement: they all require information about the total missing momentum. Unfortunately, a precise measurement of the missing momentum is often difficult, for instance due to the relatively poor reconstruction of the jets that are usually a part of the overall event structure. This is an unpleasant feature of missing momentum measurements, especially in those cases in which many of the jets that are involved in the measurement of the missing momentum are actually {\it not} involved in the decay process of interest.
%%%
Said another way, 
in general, the missing transverse momentum is measured as the opposite of the sum of the momentum of all the reconstructed objects (leptons, jets, photons,...) in the event, which means that the measurement of the missing momentum is an inherently ``global'' measurement of said event. 

In light of this, we have recently proposed complementary methods for mass measurements which instead use {\em only} the 
energy of the visible particles. The reason to pursue this strategy is, of course, that it relies intrinsically on more ``local'' information, ideally using only a subset of the particles coming from a given decay chain.\footnote{See also Refs.~\cite{Cheng:2007xv, Cheng:2008mg, Cheng:2009rt, Han:2009ss, Han:2012nr, Cho:2012gd} for other recent methods of mass measurement that do not use the missing transverse momentum and Ref.~\cite{Barr:2010hs,Gripaios:2011kk} for a general review of mass measurement methods.}
The main idea behind the method that we propose is to use the energy spectra of the visible particles. 
The basic result upon which the method is predicated was shown in Ref.~\cite{Agashe:2012bn}: namely, for a mass{\it less} child from the {\it two}-body decay of an {\it un}polarized parent, the peak in the energy spectrum of the child (henceforth denoted as ``energy-peak'') seen in the laboratory frame (henceforth denoted by ``laboratory-frame'' energy)
is the same as the fixed value of its energy
in the rest frame of the associated parent (henceforth denoted by ``rest-frame'' energy). The latter value is given by a simple relation in terms of the masses of the two massive particles (the parent and the {\em other} child) involved in the decay, and hence can give information about these masses.
In a subsequent paper~\cite{Agashe:2013eba}, we then applied this observation to measuring the unknown masses in the semi-invisible decay of a heavy new particle involving a multi-step cascade of {\em two}-body decays.\footnote{We also showed that our energy-peak result of Ref.~\cite{Agashe:2012bn} can be used for ``counting"' DM particles in decays~\cite{Agashe:2012fs}, which is a powerful probe of the DM stabilization symmetry.}

In this paper, we consider instead a {\it single}-step, {\em three}-body, semi-invisible decay of a heavier new particle, which we denote as
\bea
B \rightarrow A a b \label{eq:decayInt}
\eea
where $a$, $b$ are visible SM particles and $A,\;B$ are massive new particles with $A$ assumed invisible. In order to deal with this specific decay topology, we need to extend the result of Ref.~\cite{Agashe:2012bn} to multi-body decays.  The key idea is to map a multi-body final state into a two-body one, by the factorization of phase-space. In carrying out this mapping, we will take particular care in correctly partitioning and grouping the multi-body decay products and selecting an appropriate region of the multi-body phase space upon which to apply the above two-body result.

\begin{figure}
\centering
\includegraphics[scale=0.8]{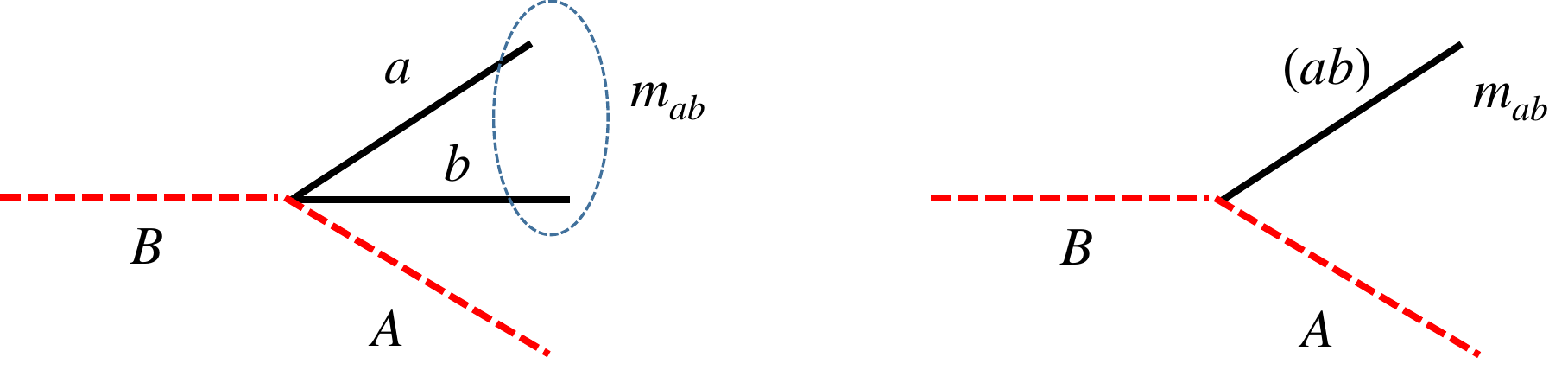}
\caption{\label{fig:EffectiveSystem} The three-body decay of interest (left panel) and the effective two-body decay (right panel) with the mass of the visible system being $m_{ab}$. $(ab)$ denotes the effective visible pseudo-particle formed by the two visible particles $a$ and $b$.}
\end{figure}

To reduce the multi-body final state to a two body one, we first form a compound system made of all of the visible particles, labeled as $a$ and $b$  in eq.~(\ref{eq:decayInt}). We denote this compound system as $(ab)$ and graphically represent the corresponding partitioning in Figure~\ref{fig:EffectiveSystem}. 
After this partitioning the decay does not yet look like a truly two-body decay because the compound system $(ab)$ does not have a fixed mass. The combination of $(ab)$ will have its own phase space in invariant mass.
This is apparent from the well-known~\cite{pdg} recursive formula for the multi-particle phase-space of $N$ particles of masses $m_{1}, ..., m_{N}$, which can be thought of as the sum of many two-body phase-spaces, a single particle on its own and the remaining $(N-1)$ particles clustered into a single object whose mass now depends on the momenta and the angles between the $N-1$ clustered particles: 

\bea
\phi_{N}(m_{1},... ,m_{N})=\int  d\mu \, d\phi_{2}\left(m_{N},\mu(m_{1},...,m_{N-1}) \cdot d\phi_{N-1}(m_{1},...,m_{N-1})\right).
\eea
Considering each value of the masses that the compound system $(ab)$ can take separately, we can regard the $N$ body final state as a weighted sum of a collection of two-body systems, each of which is characterized by the mass of the compound system denoted by $\mu$ and its probability $d\phi_{N-1}$. This probability, together with the actual squared matrix element of the decay, would give the rate of decay in that particular kinematic configuration. In the following, we do not assume any knowledge of the matrix element of the decay and we shall make no use of these rates; all we will need for our strategy to work is the ability to represent the multi-body final state as the sum of the collection of all possible two-body final states. The fact that we do not need to know the rate for each possible kinematic configuration of the multi-particle final states is a remarkable point of strength of our method; it is especially powerful when applied to newly discovered particles, as their matrix elements are {\em a priori} essentially unknown.

For the case of a three-body decay, the above outlined procedure gives 
     $$B\to Aab = \sum_{m_{ab}} \, \left( B \to A \, (ab)_{m_{ab}}\right) \,,$$
     where the equality should be taken in the sense of an equivalence. We also remark that for practical reasons the integral for the phase-space factorization formula has been discretized. In this way, we can form a finite number of compound systems $(ab)_{m_{ab}}$ of mass $m_{ab}\pm\delta$ with $\delta \ll m_{ab}$. This procedure ensures that each of the compound systems has an approximately fixed invariant mass, and we can think of it as a ``pseudo-particle'' having a width of order $\delta$. This means that we partition, or ``slice'', the data according to the total invariant mass of a compound particle formed from the $a$ and $b$ particles and apply the result for two-body decays to each mass partition as the overall system has been reduced to an effective two-body decay. In the rest frame of the parent particle, each partition of the pseudo-particle $(ab)$ has energy: %%
\bea
   E^{*}_{(ab)}=\frac{m_{B}^{2}-m_{A}^{2}+m_{ab}^{2}}{2m_{B}}\; ,   \label{eq:massiveestarrest}
\eea
based simply on two-body kinematics for decay of $B$ into $A$ and $(ab)$. 
Using the appropriate extension of the result in Ref.~\cite{Agashe:2012bn} to the case $m_{ab}\neq 0$  (see more on this point 
in Section~\ref{sec:massivetemplate}), we are able to extract $E^*_{(ab)}$ from the laboratory-frame energy distribution of that particular $(ab)$ compound particle. We then repeat this procedure for each of the mass partitions in the overall range of $m_{ab}$.
When plotted versus $m_{ab}^2$, the fitted data for $E_{(ab)}^*$ extracted from the energy distributions should lie along a straight line as per eq.~(\ref{eq:massiveestarrest}).
It is straightforward to see that $m_B$ can be determined from the slope of this line and that $m_A$ can be determined from the intercept on the vertical axis once $m_B$ has been determined. The available information can be fully utilized in constructing this straight line by analyzing the data in all slices.\footnote{In practice, one could end up not using some of the slices if treating them becomes too problematic, e.g. because of backgrounds or sensitivity to the cuts. The fraction of unused slices will in any case be kept to a minimum.}

We remark that, although the characteristic signature 
of the production of an invisible particle $A$ is missing momentum, our method does not make explicit use this quantity and yet still offers a way to obtain a measurement of mass of this particle!
In other words, any specific property of the invisible particle is almost completely irrelevant to our method, except for the assumption that there is at least one invisible particle per decay chain. We again emphasize that this achievement is remarkable, especially in comparison with other mass measurement methods involving $\misse$ such as $M_{T2}$ and its variants.

Despite the simplicity of the general idea, there are still some potential issues that should be properly dealt with in order to successfully execute the strategy outlined above.
First, the compound particle $(ab)$, the visible child particle from the {\it effective} two-body decay, has a non-negligible mass; thus, it is essential to generalize the result in Ref.~\cite{Agashe:2012bn} on the energy-peak to the case for a {\it massive} visible child particle. 
Our companion paper~\cite{Agashe:2015ike} will be devoted to studying how to deal with these massive child particles in more detail. In the present work, we shall merely report the final result of this companion paper and use this result for our present investigation. Nevertheless, the discussion presented here is largely independent of the derivation of this result.

As mentioned earlier, the DM model under consideration has the parent particles being produced {\em in pairs}. If both of them decay to the same final state, a combinatorial ambiguity arises in attempting to correctly partition and group those particles originating from the same parent; multiple pairs can be formed from the final state as seen in the detector, but it is not known \emph{a priori} which is the correct pairing - that is, that the particles in the pairing originated from the same decay. This partitioning is a crucial necessity in forming the $(ab)$ compound system that plays the role of the child {\it pseudo}-particle. 
%%%%
%
For this reason, we allot  Section~\ref{sec:eventmixing}  to the thorough discussion of the treatment of this combinatorial ambiguity. In particular we propose an ``event mixing'' technique~\cite{Albrow:1976jm} and Refs.\cite{Drijard:1984pe,Kjaer:2001ay,CMS-Collaboration:2013lr, Dutta:2010qm, Dutta:2011uq} as a way to remove the combinatorial background.

%%%
%
%

To illustrate our method we discuss in detail its application to a specific process. As a concrete example, we choose the pair production of gluinos in an R-parity conserving supersymmetric model. Here the gluinos are assumed to subsequently decay into $b \bar{b}$ and an invisible light stable neutralino via an {\em off}-shell bottom squark:
\bea
pp\to \tilde{g}\tilde{g} \to  b\bar{b}b\bar{b}\chi\chi.
\eea 
This scenario is chosen primarly because it has been thoroughly studied in the literature, and thus should be familiar and interesting to a large audience. Indeed, this process has also been investigated at the LHC~\cite{CMS-PAS-SUS-13-019,CMS-Collaboration:2013pd,CMS-Collaboration:2013bo,ATLAS-CONF-2013-061,CMS-PAS-SUS-14-011,ATLAS-Collaboration:2014ud}. In order to provide a fully realistic example, and to demonstrate some of the issues that arise in using our method, we shall incorporate the relevant Standard Model backgrounds in our analysis as well.
 We take particular care in devising cuts for background rejection so that these selections do not affect the shape of the energy spectrum near the peak, which is the critical region of interest for our energy peak method. Obviously, the optimization of these cuts is a process-dependent issue, and so must be evaluated on a case-by-case basis. The goal of our discussion is to present potential systematic uncertainties and biases arising from the specific details of our method, such as those induced by phase space slicing, event mixing, imperfect knowledge of the background, and overly restrictive event selection criteria. We also present other complementary observables that enhance the findings obtained using the energy peak method, one of which is the kinematic {\it endpoint} of the di-jet invariant mass distribution.

The rest of the paper is organized as follows. We continue in the next section with a discussion of a template function used to describe the energy spectrum of a massive child particle. Section~\ref{sec:eventmixing} is devoted to dealing with the combinatorial ambiguity inherent in our chosen decay topology. Then in Section \ref{sec:example},
we detail our selected signal process and the relevant backgrounds, both those from SM processes and the ``event mixing'' scheme for signal combinatorics. Section~\ref{sec:results} contains the main results for the mass measurement of the aforementioned example process together with a discussion of several opportunities for improvement to the method. In Section~\ref{conclude} we present our conclusions and outlook.

\section{A template for the energy spectrum of a massive child particle \label{sec:massivetemplate}}
 As outlined in the previous section, the essence of our mass measurement technique is to fit the data to get the value of $E^*$ for each of the fixed masses of the compound system $(ab)$ and fit them onto the straight line in eq.~(\ref{eq:massiveestarrest}). The mass of the compound system $(ab)$, being a system of two particles, is not fixed and in general spans a range fixed by the masses of all the particles in the decay. Since we are {\it a priori} unaware of the masses of the parent and invisible child particles, it is not possible to know whether or not a given value of $m_{ab}$ is small enough in comparison to those unknown masses to justifiably trust the validity of our previous results for {\it effectively} massless child particles~\cite{Agashe:2012bn}, and thus we are motivated to extend the finding to massive child particles.

The primary difficulty in generalizing to a massive child is the potential loss of correlation between the peak location in the laboratory-frame energy distribution of the massive child and its energy in the rest frame of the parent. This can readily be seen when considering the decay of a massive particle $B \to X \, \psi$ at the kinematic end point of the phase-space, i.e., $m_{B}=m_{X}+m_{\psi}$. For any value of $m_{X}$, $\psi$ will be at rest in the $B$ rest frame, hence $E^{*}_{\psi}=m_{\psi}$. If each particular event is boosted to the laboratory frame, the energy of $\psi$ becomes simply $\gamma_{B}m_{\psi}$, with $\gamma_{B}$ being the boost of particle $B$ relative to the laboratory. This direct linear relationship between $E_{\psi}$ and $\gamma_B$ implies that the shape of the energy distribution of particle $\psi$ in the laboratory frame should simply be that of the boost distribution of particle $B$. In this case, it is clear that the peak of the energy distribution of the massive child $\psi$ carries essentially no information about the masses; rather, it carries information on the most probable boost of particle $B$. This is contrast to the ``invariance'' that holds for a massless child:  the energy-peak in the laboratory frame is the same as the rest-frame energy value irrespective of the details in the boost distribution of particle $B$.

For a more formal understanding of this problem, it is instructive to analyze the Lorentz transformation of a massive child particle from the rest frame of its parent particle, where it has energy-momentum $(E^{*},p^{*})$, to the laboratory frame. 
Given the boost factor $\gamma$ of the parent particle and the emission angle of the child $\theta^*$ relative to the boost direction, we find the energy of the child particle in the laboratory frame (denoted by $E$) to be
\bea
E=
E^{*}\gamma\left( 1 + \beta^{*}\beta \cos\theta^{*} \right) \label{eq:LabEMassiv}\,,
\eea 
where we have used $p^{*}=\beta^{*}E^{*}$. We observe that the laboratory-frame energy $E$ becomes equal to the rest-frame energy $E^*$ only if
\bea
\cos\theta^* =-\frac{1}{\beta^*\beta}\left(\frac{\gamma-1}{\gamma} \right). \label{eq:criticalangle}
\eea
Denoting this $\theta^*$ as the ``reference'' angle, we see that any value of $\cos\theta^*$ smaller (larger) than eq.~(\ref{eq:criticalangle}) gives rise to a laboratory-frame energy value $E$ smaller (larger) than $E^*$.

To ensure the existence of the reference angle, the boost factor should satisfy the following relation:
\bea
\gamma < \frac{1+(\beta^*)^2}{1-(\beta^*)^2}=2(\gamma^*)^2-1 \equiv \gamma_{\textnormal{cr}} \label{eq:criticalboost}\,.
\eea 
When this condition is satisfied, the energy distribution in the laboratory frame is non-zero at $E=E^*$, which is a necessary, but {\it not} sufficient, condition to have a maximum at $E^*$. Obviously, if for some $\gamma$ the condition set by eq.(\ref{eq:criticalboost}) is not satisfied, then $E>E^*$ for all $\cos\theta^*$, which potentially invalidates the statement that the peak in the laboratory-frame energy distribution appears at $E^*$. Actually, for the typical boost distributions of parent particles produced at hadron colliders, one can see that if any of the boosts of the parent particle(s) lie outside of the range given by~(\ref{eq:criticalboost}), it is then guaranteed that the peak of the energy distribution in the laboratory frame will {\it not} be located at the rest-frame energy value~\footnote{The displacement of the maximum with respect to $E^{*}$ may still be small, but strictly speaking, the ``invariance'' that we demonstrated in Ref.~\cite{Agashe:2012bn} for the massless particle will be broken.}. A more rigorous method of characterizing the energy distribution of a massive child will be presented in a separate work~\cite{Agashe:2015ike}. We hope that the argument above, while not as rigorous as our companion paper, will convince the reader that the maximum boost of the parent particle is the key parameter that controls the position of the peak in the laboratory-frame energy distribution of a massive child particle.

On top of affecting the peak position, the overall shape of the energy distribution for the massive child is expected to differ from that for the massless child. This means that the function used to fit the massless child energy spectra in previous works cannot be used in the present work. In order to obtain a suitable description of the massive child energy spectrum, we revisit the corresponding discussion for the case with a massless child particle. The value of the energy distribution
at a given laboratory-frame energy $E$ is given by a Lebesque-type integral within the range of $\gamma$ values contributing to the $E$ together with the associated weight for the $\gamma$~\cite{Agashe:2012bn}. For the case at hand, this range is found by solving eq.~(\ref{eq:LabEMassiv}) for $\gamma$ with $\cos\theta^*=\pm 1$, which gives: 
\bea
\gamma_+(E)&\equiv&\gamma^{*2}\left(\sqrt{1-\frac{1}{\gamma^{*2}}}\sqrt{\frac{E^2}{E^{*2}}-\frac{1}{\gamma^{*2}}}+\frac{E}{E^*} \right) \label{eq:gammap}\,, \\
\gamma_-(E)&\equiv&\gamma^{*2}\frac{E}{E^*}\left(1-\sqrt{1-\frac{1}{\gamma^{*2}}}\sqrt{1-\frac{E^{*2}}{\gamma^{*2}E^2}} \right). \label{eq:gammam}
\eea 
We see that for a massless child particle, i.e., $\gamma^{*}\to \infty$, $\gamma_{+}(E)$ diverges, whereas in the same limit $\gamma_-(E)$ converges to a finite value: 
\beq
\lim_{\gamma^{*}\to \infty} \gamma_{-}(E) = \frac{E}{E^{*}}+\frac{E^{*}}{E} \equiv \gamma_{-}^{(\infty)}(E) \label{eq:gammaminfty}\,.
\eeq
In light of eq.~(\ref{eq:gammaminfty}), we can express the energy spectrum for a massless child that we used in previous works~\cite{Agashe:2012bn} as
\bea 
\label{eq:masslessTemplate}\exp\left( -w\cdot \gamma_{-}^{(\infty)}(E) \right).
\eea 

Motivated by the success of this exponential form in the massless case \cite{Agashe:2012bn,Agashe:2013eba} and exploiting the identification in the massless limit eq.
(\ref{eq:masslessTemplate}), we propose the following ansatz on the shape of the laboratory frame energy spectrum of a massive child 
\bea
f(E)=N\left(\exp[-w\cdot\gamma_-(E)]-\exp[-w\cdot\gamma_+(E)]\right)\,,\label{eq:massiveTemplate}
\eea
where $N$ and $w$ are a normalization factor and the width of the function, respectively.
A complete evaluation of the accuracy with which this function describes the energy spectrum in the laboratory frame of a massive child will be presented in a dedicated work~\cite{Agashe:2015ike}. For the purposes of this paper, it will be sufficient to know that this function reproduces our ansatz for massless child particles for $\gamma^{*}\to \infty$. In any case, we will explicitly show that this function provides a good description of simulation data for our example process below.  

A comment on the location of the maximum of this function is in order. The maximum of this function coincides with $E^{*}$ only in the limit $w\to \infty$, in which the function becomes a $\delta$ function. For all finite values of $w$, the actual location of the maximum of the function is slightly larger than $E^{*}$. However, we empirically observe that for parent particles that would typically be produced at colliders, and for $\gamma^{*}$ somewhat larger than 1, the typical value of $w$ is large enough that this effect is negligible. Therefore, we expect that eq.~(\ref{eq:massiveTemplate}) properly describes a large class of energy spectra.
 Because the peak location and the $E^*$ parameter are no longer necessarily the same \footnote{Again, for the case with massless children, $E^*$ conforms to the peak irrespective of $w$. }, the determinations of the best fit values of $E^*$ and $w$ are interrelated when fitting the data to the massive template function eq.(\ref{eq:massiveTemplate}). 
The fact that the maximum of the spectrum is a function of both $E^*$ and $w$ is an inherent feature of our ansatz for the massive child energy spectrum, which did not exist in the massless case of Ref.~\cite{Agashe:2012bn}. In this paper we will study the possible effects that arise in our explicit example due to this feature of eq.(\ref{eq:massiveTemplate}).  For a fully general investigation of this issue we refer to our companion paper~\cite{Agashe:2015ike}.
In Sec.~\ref{sec:example} we shall fit the energy distribution of each mass partition of the pseudo-particle $(ab)$ both with the new template for massive children  eq.~(\ref{eq:massiveTemplate}), and with its massless limit (i.e $\gamma^{*}\to \infty$), the latter of which was the template employed previously for massless child particles. The comparison of the results from these two templates will allow us to demonstrate the necessity of using eq.~(\ref{eq:massiveTemplate}) and generalizing what we had used in our previous work~\cite{Agashe:2012bn,Agashe:2013eba}.

\section{ ``Event mixing'' to estimate the combinatorial background \label{sec:eventmixing} }

As explained in the Introduction and depicted in Figure~\ref{fig:EffectiveSystem}, the success of our strategy hinges on correctly identifying the pairs of particles coming from the same decay side. If this identification is done correctly, the idea of phase-space factorization can safely be applied to reduce the multi-body final state to a two-body final state. Generally speaking, the identification of the correct pairs of particles to be grouped together is a tremendously difficult task, as we have no systematic way of knowing which particles are related to the same decay, and can thus be correctly paired in the analysis.

One approach to surmounting this challenge is to attempt to correctly identify the pairs of particles that come from the same decay by exploiting some preferential kinematic correlation between particles that originate from the same decay. This correlation could then be translated into a selection criterion that would correctly pair the appropriate particles with relatively high accuracy. Of course, this selection process would not be $100\%$ effective and would fail to pick the correct pairings for some fraction of events. As a result, we would be left with a certain amount of combinatoric pollution from pairings whose constituent particles did not come from the same decay. Several event-by-event strategies have been developed to identify which pairs of particles come from the same decay (see, for example, Refs.~\cite{Rajaraman:2010hy, Bai:2010hd, Blanke:2010cm, Baringer:2011nh, Chatrchyan:2013boa}). It is however generally true that, in order to maximize the chance of pairing particles correctly, the kinematic selection criteria which form the basis of each method must be rather restrictive, so as to guarantee a sufficient rejection of unwanted pairings. Events that pass these highly constrained kinematic criteria will be preferentially selected from isolated regions of the kinematic phase space of the scattering, and therefore, the kinematic distributions of the final state will be significantly altered by the imposition of these criteria. Our method of mass extraction relies critically on the fidelity of the energy distribution around the location of the peak; without this, the templates we use to extract the masses return biased information. Thus, because the above set of procedures for selecting correct decay siblings greatly disturbs the resultant kinematic distributions, they are not suitable for use alongside our method. For this reason, we do not even attempt to identify the correct pairings in each event; we instead try to obtain the energy {\it distribution} of the correct pairs without knowing which are the correct pairs event-by-event.

In order to determine the {\it distribution} of a given observable as it would arise from  picking just the correct pairs, we use an event mixing technique whose basic idea is as follows. We consider a scattering 
\bea
pp \rightarrow B_1 B_2 \rightarrow (A_1 a_1 b_1)(A_2 a_2 b_2),
\eea
where two heavy particles $B_{1}$ and $B_{2}$ are produced and decay to final states of the same kind, $B\to A a b$, which are labeled to correspond with their respective parent.\footnote{Strictly speaking, $A_i$'s need not be invisible as long as they are distinguishable from $a_i$ and $b_i$, which are assumed indistinguishable from each other.} In the analysis of events of this nature, we follow the procedure laid out in Section~\ref{sec:generalstrategies} for making pseudo-particles out of the $a$ and $b$ particles for \emph{all possible equivalent pairings} (e.g. $a_1$ with $b_1$, but also $a_1$ with $b_2$ and so on.) 
From these pairs, we obtain a fully inclusive distribution of the observable of interest, which in our case here is the total energy of the pair, 
\bea
\frac{d\sigma}{dE_{ab}}(\textrm{all~pairs}).
\eea

In order to obtain the distribution stemming only from the pairs of particles coming from the same decay, it is sufficient to come up with an estimate of the distribution that stems from the pairs that we would like to discard and subtract it from the fully inclusive distribution: 
\bea
\frac{d\sigma}{dE_{ab}}(\textrm{same-decay~pairs}) = \frac{d\sigma}{dE_{ab}}(\textrm{all~pairs}) - \frac{d\sigma}{dE_{ab}}(\textrm{different-decay~pairs}).
\eea
This equality might seem trivial, but is in fact very powerful. To wit, the object that we must have for our method to be successful - the distribution of pairings from the same decay - which is quite difficult to obtain in itself, is now expressed in terms of two objects that are much simpler to obtain. The first piece is obviously attainable, as it is simply the distribution from all the pairs that can be formed in each event. The second piece, the distribution from pairs {\it not} coming from the same decay, can be estimated by the ``event mixing'' technique, which is based on making a distribution from pairs of jets that come from \emph{different events}. 
The intuition that justifies the usefulness of the event mixing technique originates from the fact that pairs $(a_{1}b_{2})$ and $(a_{2}b_{1})$ from the same event are made of particles which are produced with almost no kinematic correlation. Therefore, it seems reasonable to mimic the effect of these ``incoherent'' pairs with the pairs of particles taken from different events, which intuitively have no correlation. 
More precisely, we can see that the phase-space point from which $a_{1}$ and $b_{1}$ originate in the decay of $B_{1}$ is very close to being uncorrelated with the phase-space point from which $a_{2}$ and $b_{2}$ originate in the decay of $B_{2}$. This approximately vanishing correlation between the products of \emph{different} decays implies that pairs formed by particle $a_{1}$ taken from one event, and particle $b_{2}$ taken from another event are expected to be  statistically  equivalent  to pairs $(a_{1}b_{2})$, where both particles are taken from the same event. This means that the {\it distributions} of a quantity over a given sample of events obtained from either the $(a_{1}b_{2})$-type incorrect pairings within the same event or from pairing $a_{1}$ in one event and $b_{2}$ in another event are equivalent. This implies that we can estimate the distribution stemming from pairs of particles from the same decay as  
\beq
\label{eq:eventmixingsubtraction}
\frac{d\sigma}{dE_{ab}}(\textrm{same-decay~pairs}) \simeq \frac{d\sigma}{dE_{ab}}(\textrm{all~pairs}) - \frac{d\sigma}{dE_{ab}}(\textrm{different-events~pairs})\,. 
\eeq
This observation is at the center of the event mixing idea, and we henceforth denote the procedure described by the right-hand side of eq.~(\ref{eq:eventmixingsubtraction}) as ``mixed event subtraction''.

The plausibility of the distribution obtained by pairing particles from different decays within the same event being equivalent to the distribution arising from pairings between different events can be seen intuitively in certain simple cases. To illustrate, imagine having an ensemble $E$ of pairs of particles $(B_{1},B_{2})$, 
\bea
E=\left\lbrace (B_{1}^{(1)},B_{2}^{(1)}), (B_{1}^{(2)},B_{2}^{(2)}), (B_{1}^{(3)},B_{2}^{(3)}), ...  \right\rbrace,
\eea
 where the superscripts denote the associated event number. For simplicity, we take particles $B$ to be scalars sitting at rest in the laboratory, where they decay $B\to abA$.  It is obvious that the distributions made from pairs coming from different $B$ particles in the {\it same} event of the ensemble are the same as the distribution made from pairs coming from $B$ particles from {\it different} events in the ensemble.  In this example, we are simply sampling the phase space of the $B$ decay in two different ways, in one case taking kinematic information from instances of the decay that happen at the same time and in the other taking that information from instances of the decay that are separated in time. 

However, one must note that the situation described above may not correspond to reality. As a concrete example, we take the pair of $(B_1^{(i)}, B_2^{(i)})$ and assume that the system of the two $B$ particles has a center of mass energy $\sqrt{\hat{s}_i}(>2m_{B})$. In the laboratory frame, the phase-space accessible to the decay products of the two $B$ particles depends on $\sqrt{\hat{s}_i}$. If we consider the $j$th event, with the intention of mixing particles from the $i$th and $j$th events, we are forced to confront the fact that the center of mass energy in the $j$th event, $\sqrt{\hat{s}_j}$,  and thus the phase space accessible to the $j$th event's particles, is different than that of $i$th event. The mismatch in the phase-space accessible to particles in events at different $\sqrt{\hat{s}}$ is clearly a potential source of error in the identification of the ``different-decay'' and ``different-event'' distributions, which naturally poses an a threat to the successful application of the event mixing technique.

It is quite difficult to estimate the size of this inaccuracy, though we expect that for typical situations at hadron colliders the event mixing technique works quite well. One reason for this is because of the small variance of the boosts of the $B$ particles produced in typical collisions at hadron colliders. In addition to this effect, however, there may be other potential sources of error in the event mixing technique, and a case-by-case study is needed to check the performance of the method. Because of this, we take a pragmatic approach in the following and apply the event mixing technique to our example while explicitly checking the performance of this method for our example process. 

\section{Application to the gluino decay \label{sec:example}}
We now demonstrate how the general strategy detailed above is realized by taking as an example a particular gluino decay channel. We first illustrate the signal process and its particular characteristics, and then move to discussing the possible backgrounds of this signal process. The discussion of these backgrounds is separated into two categories: 1) the real background from SM processes, and 2) the systematic background from incorrect pairings of the final state particles used in forming the invariant mass and energy distributions. 
As we detail the various processes involved, we shall utilize Monte Carlo simulation to generate the relevant event samples, construct and analyze the appropriate kinematic data from these samples, and end with a discussion of the effectiveness of our technique based on the results of this analysis.

\subsection{ Signal process: gluino decay \label{sec:signaldetails}}
We apply the general idea developed in the previous sections for the case of pair-produced SUSY gluinos and their subsequent decay into two bottom quarks and a neutralino via a three-body decay:
\bea
p p \rightarrow \tilde{g} \tilde{g} \rightarrow \bar{b}b\bar{b}b +\chi\chi \label{eq:sigprocess}
\eea
at the 14 TeV LHC. In terms of the notation used in Section\ref{sec:generalstrategies}, the gluino and the neutralino correspond to particles $B$ and $A$, respectively, and two visible particles $a$ and $b$ are the  bottom quark and anti-quark in a decay chain. In reality, the particle detector cannot reliably discriminate between bottom and anti-bottom, thus particles $a$ and $b$ in this example are considered {\it in}distinguishable.

Though we are using the specific decay above as a concrete example, we emphasize that the decay mode and underlying model at hand are chosen only to enable us to demonstrate the proposed technique and that the general idea can be applied to multi-body decay processes in other models.  We also point out that the applicability of the method is not affected in any way by the strengthening of bounds on supersymmetric particles, because the method is applicable for parent and invisible (child) particles of any mass. To illustrate our technique, we choose the masses of these particles to be 
$$m_{\tilde{g}} = 1.2 \tev ,\quad m_{\chi}=100\gev$$
with a decoupled bottom squark, and assume that the only decay mode of the gluino is a three-body decay in the form of $b\bar{b}\chi$~\footnote{During the completion of this work the limit on the gluino mass given by the LHC experiments has risen to about 1.4 TeV for light $\chi$~\cite{CMS-PAS-SUS-14-011,ATLAS-Collaboration:2014ud}. Despite these new limits ruling out the spectrum we consider at a 95\% confidence level, this spectrum still serves its purpose as an illustration of the technique. It should be remarked that we do not expect qualitative differences in the application of our strategy to the mass measurement of a heavier but not yet experimentally excluded gluino. }. 
The Monte Carlo signal for our study is simulated using \texttt{MadGraph5}~v1.4.8~\cite{Alwall:2011uj} and the structure of the proton is parametrized by the parton distribution functions (PDFs) \texttt{CTEQ6l1}~\cite{Pumplin:2002vw}, evaluated with the default renormalization and factorization scale settings of \texttt{MadGraph5}. The production cross section of the paired gluinos is computed with \texttt{MadGraph5} and is reported in the first column of Table~\ref{tab:crossX}.  Since we assume that all produced gluinos decay into $b\bar{b}\chi$ as described above, $\sigma(pp\rightarrow \tilde{g}\tilde{g})$ is  equal to $\sigma(pp\rightarrow \tilde{g}\tilde{g}\rightarrow b\bar{b}b\bar{b}\chi\chi)$. 

\begin{table}[t]
\centering
\begin{tabular}{c||c|c||c|c}
\hline
 & $4b+\misse$  & $4b+\misse$ & $4b+Z(\rightarrow \nu\bar{\nu})$ & $t\bar{t}b\bar{b}$ \\
 & (before cuts) & (after cuts) & (after cuts) & (after cuts) \\ 
 \hline
 $\sigma$ [$fb$] & 54.74 & 36.53 & 0.48 & 0.15 \\
 \hline
\end{tabular}
\caption{\label{tab:crossX} The cross sections for the signal process (before and after a set of cuts are imposed) and the (main) background processes (only after cuts are imposed). The cuts are described in eqs.~(\ref{eq:cutforb})-(\ref{eq:delphicut}), and also include all the identification and the isolation criteria explained in the text. The effect of the $b$-tagging efficiency is not taken into account by the numbers in this table.}
\end{table}

The neutralinos in the final state of our signal do not interact with the detectors of the LHC, resulting in a missing transverse momentum.
The four bottom quarks give rise to jets of hadrons - particularly, $B$-hadrons. The particular characterisitcs of the $B$-hadrons in the jets allows us to distinguish this type of jet from other jets that do not originate from the bottom quark, and it is possible to see the traces of $b$-quark-initiated jets and tag them in a large fraction of the events. With the requirement that four of the reconstructed jets in the final state have this tag, the signal will feature four bottom jets plus missing transverse momentum, $4b+\misse$.

Before closing this section, we remark that the chosen example process poses an extra challenge in the application of our method. In fact all visible final state particles are {\it in}distinguishable, hence there are three different  ways to form {\it pseudo}-particles from these $b$-tagged final state particles that must be all be considered in the analysis. Together with the SM backgrounds, this can be interpreted as another background, as explained in the next section.

\subsection{Backgrounds}

In this section, we discuss the backgrounds relevant to the signal process defined in the preceding section. As mentioned earlier, there are two types of backgrounds: those coming from Standard Model processes which give rise to the same signature as our supersymmetric process,  and that which comes from taking the wrong pairings of two visible particles when we evaluate both the energy sum and the pair-wise invariant mass. We start by discussing  the ``real'' background from the Standard Model and then we discuss the combinatorial background.

\subsubsection{Standard Model backgrounds and event selection \label{sec:eventselection}}
For our collider signature $4b+\misse$, the following two processes in the SM are identified as the major backgrounds:
\bea
pp \rightarrow bb\bar{b}\bar{b} +Z \rightarrow bb\bar{b}\bar{b}+\nu\bar{\nu} \;\;\;\quad\quad\hbox{and}\;\;\;\quad\quad pp\rightarrow t\bar{t}b\bar{b}\,. \nonumber
\eea
The Monte Carlo generation of background events is done using the same event generator and input PDFs as those for the signal events.
Since the detector signature from these interactions is exactly the same as the one used in our earlier work Ref.~\cite{Agashe:2013eba}, we adopt a similar strategy for handling these backgrounds with only slight modifications. The $Z$ boson background is irreducible, whereas the $t\bar{t}b\bar{b}$ is reducible and can be reduced so that it becomes sub-dominant with respect to the $Z$ boson background. The $t\bar{t}b\bar{b}$ background might seem different from the signal process in terms of its partonic final state, but it can mimic our signal, and thus become a relevant background, by ``losing'' some of the final state partons in the detectors. In order to match with the signal's detector signature, the two $W$  bosons originating from the decay of the two top quarks must go undetected, which makes those $W$ bosons the main source of $\misse$ in the $t\bar{t}b\bar{b}$ background. Although the rate of the detector missing the $W$ bosons is expected to be small, the sizable production rate of $pp\rightarrow t\bar{t}b\bar{b}$ can compensate for this, thus making the $pp\to t\bar{t}b\bar{b}$ process a possibly important background.

A $W$ boson will go unseen in the detector for primarily two reasons:
1) when its decay products are not within the experimental acceptance region of the detector due to having insufficient $p_T$, supernumerary $\eta$, or both, and 2)  when its decay products are not adequately isolated from other particles, i.e., they are merged with other particles in the reconstruction of a given event. For the first case, we define as missed any object that satisfies the following criteria:
\begin{itemize}
\item for jets, $p_{T,j}<30$ GeV or $|\eta_j|>5$,
\item for leptons, $p_{T,l}<10$ GeV or $|\eta_l|>3$ with $l=e,\mu,\tau$. 
\end{itemize}
In the second case, the following rules determine when a particle is missed:
\begin{itemize}
\item for merging jets, $\Delta R_{j_1 j_2}<0.4$ with $j_1$ and $j_2$ denoting any jet pairs including $b$-jets,
\item for merging leptons, $\Delta R_{jl}<0.3$ with $j$ and $l$ denoting a jet and a lepton, respectively. 
\end{itemize}
With the acceptance and isolation requirements listed above, we observe that most of the background events are from either the fully leptonic or the semi-leptonic decay channels of top quark pairs because these channels require that fewer erstwhile visible partons be missed in comparison to the fully hadronic top decay channel. 

To devise an adequate strategy for rejecting a large number of these background events while preferentially keeping the signal events, we must adopt event selection criteria that incorporate the kinematic differences between signal and background event. We observe first that, thanks to the heaviness of the parent particles, the signal events will be composed of jets that typically have a larger transverse momentum than those found in background events. This is a strong hint as to which cuts will be suitable in rejecting the background. However, one must be especially cautious in selecting these cuts 
because the method proposed here is based on extracting $E_{bb}^*$, which relies in part on the shape analysis of the energy distributions for each invariant mass slice. Therefore, cuts should be chosen such that they do not considerably distort the energy distributions. For this reason, we prefer using softer cuts than in most searches performed at the LHC, and we choose as our baseline selection criteria 
\bea
p_{T,b}> 30 \hbox{ GeV},\; |\eta_b|< 5, \; \Delta R_{bb}>0.4, \label{eq:cutforb}
\eea
for identifying the bottom jets in all events that we analyze. 
In addition, as a good approximation, a flat selection rate of 66\% is assumed for any single bottom jet to account for the $b$-tagging efficiency.\footnote{Note that in practice, this efficiency will have a dependence on $p_T$ of the $b$-jet,
thus modifying the {\em shape} (i.e., not just the normalization) of the energy spectra. Nonetheless, knowing this functional form
of the $b$-tagging efficiency, 
we can still recover the ``true" shape of the $b$-jet energy spectra.}

In order to further suppress the backgrounds from Standard Model processes, we consider requiring that events have a large missing transverse momentum. 
In signal events, the missing transverse momentum is expected to be determined by some combination of the new particle masses, and thus will be large.  On the other hand, the missing transverse momentum in background events is determined by the larger of the total hardness of the event, the mass of the $Z$ boson, or the mass of the top quark. Therefore, a large $\misse$ cut allows us to efficiently discriminate the signal events from the background. However, in our case special care is needed in deciding the scale of this cut,  as there is the risk of this cut introducing unwanted bias in the energy distributions. In particular, the missing transverse momentum can be interpreted as the recoil of the invisible particles against the visible, which seems to imply that a large $\misse$ cut is likely to select only events with very hard visible particles and correspondingly induce some bias in the $b$-jet energy spectrum toward higher energies. As mentioned before, this could lead to a misidentification the value of $E_{bb}^*$, and as a consequence, an innacurate measurement of the associated masses. Fortunately, the relatively large mass hierarchy between the gluino and the neutralino in our signal process ensures multiple hard $b$-jets on average and thus a sizable recoil for the invisible neutralinos. We therefore anticipate that the $E_{bb}$ distribution will only be mildly affected, even with a fairly hard $\misse$ cut. For our signal and backgrounds, we impose  
\bea
\misse > 200 \hbox{ GeV}\,, \label{eq:metcut}
\eea
 which strongly suppresses the backgrounds with negligible deformation of the $E_{bb}$ distributions. 
 
In addition to the $\misse$ cut, we introduce another cut that requires each $b$-jet $\vec{p}_{T}$ to have some minimum angular separation from the $\misse$ vector. 
This enables us to avoid events where the measured missing energy is caused by the mismeasurement of jets. For our analysis we require
\bea
\Delta \phi(\missev,\;\vec{p}_T^b)>0.2, \label{eq:delphicut}
\eea
 which has negligible effect on the shape of the $E_{bb}$ distributions.%
  
In Table~\ref{tab:crossX} we show the cross sections for both signal and background events after applying the set of cuts listed above.  We clearly see that the $t\bar{t}b\bar{b}$ background is sub-dominant with respect to the $Z+bbbb$ background. We also remark that the expected signal-to-background ratio ($S/B$) is  large, which is certainly favorable for extracting $E_{bb}^*$ from the $E_{bb}$ distribution. Indeed, if new physics particles are discovered in the forthcoming runs of the LHC, it would then be natural to discuss measuring their masses in the channels where there is a clean signal, and hence a large $S/B$. In this sense, the context in which we present our mass measurement technique is expected to be typical when attempting a mass measurement beyond the precision of the order of magnitude, as we do here.

Other than the above-mentioned backgrounds, QCD multi-jet production  $pp\rightarrow bb\bar{b}\bar{b}$ is another possible source of background events from the SM, in which the missing transverse momentum typically arises from imperfectly measuring the energy of jets. Unsurprisingly, an accurate estimation of this background is quite challenging because it involves detector effects. We expect that a great deal of the QCD multi-jet background would largely be suppressed by the cuts eqs.~(\ref{eq:cutforb}),~(\ref{eq:metcut}) and~(\ref{eq:delphicut}) to the point that it becomes sub-dominant, and in the following we do not taking this background into account (see, for example, Ref.~\cite{Agashe:2013eba}).

\subsubsection{Combinatorial background and mixed event subtraction \label{sec:MESS}}

As mentioned earlier, the procedure of phase space slicing inevitably requires the formation of the invariant mass of two visible objects. Since the process eq.~(\ref{eq:sigprocess}) under consideration has pair-produced parent particles which both decay  to {\it in}distinguishable visible children, grouping the four $b$-jets into two pairs gives the correct choice in only one case out of the three possible combinations. Following the strategy outlined in Section~\ref{sec:eventmixing}, we form all possible pairs and obtain an inclusive energy distribution. We then subtract the contributions originating from the wrong combinations by estimating the corresponding distribution through the event mixing technique as described before. In order to validate the performance of this mixed event subtraction scheme in our example, we first study a large number of pure signal events where no selection cuts are imposed and without including backgrounds. We then discuss how the inclusion of backgrounds complicates the estimate of the combinatorial background when using the mixed event subtraction.  

\paragraph{i) Pure signal:} For our study it is necessary to check that $b$-jet pairs' energy and invariant mass distributions are well reproduced by the mixed event subtraction scheme.
In order to apply it as per eq.~(\ref{eq:eventmixingsubtraction}), we need to obtain the distributions of observables given by forming pairs of $b$-jets belonging to different events as explained in Section~\ref{sec:eventmixing}.  In principle, there are several options for choosing the two events that one can use to compute these observables. For example, one can compute the observables using all possible pairs of events, meaning that each event is reused many times, or one can use a procedure such that each event is used only a few times. The detailed way in which the event mixing is done can, in principle, affect the result. However, in most cases, the alternative methods give rise to only minor differences in the relevant result~\footnote{One can also form ``events'' out of randomly selected sets of four particles from the entire event sample and compute the observables by forming pairs from the particles now constituting these new ``events.'' We evaluated the efficacy of constructing the distribution to be subtracted using this alternative method and found little difference between the end results, both in the distributions and in the energy and mass values extracted.}. Among those possibilities, we show results for which the events were mixed as follows: given a sample of $N$ events,
\begin{itemize}
\item[(1)] we first randomly shuffle and reorder the events to remove any potential correlation between events arising from the way in which the events were generated, 
\item[(2)] compute the different-event observables by taking $b$-jets in the $i$th and $(i+1)$th events,\footnote{The last event is mixed with the first one.} so that each event is used twice
\item[(3)]  finally renormalize these distributions to weigh as much as the contribution from the incorrect pairings in the signal sample that we intend to remove, i.e., two thirds of the total  number of events in the signal sample.
\end{itemize}

We label the inclusive invariant mass distribution formed from all the pairs in the same event as  $F_{SE}(m_{bb})$ and we denote as $F_{DE}(m_{bb})$ the distribution obtained from pairs in different events. Then, given an invariant mass value, we take from the same-event sample all the pairs whose invariant mass lies within the range  of interest and plot the spectrum of the energy of the sum of the two $b$-jets, $E_{bb}$. We repeat the same operation in the different-event sample, and we obtain the $E_{bb}$ spectrum for the fixed ranges of $m_{bb}$. Denoting the spectrum from the same-event pairs as $f_{SE}(E_{bb})$, the spectrum from different-event pairs as $f_{DE}(E_{bb})$, and the resultant spectrum from the mixed event subtraction as $f_S(E_{bb})$, our estimate of the energy distribution from the correct pairs is:
\bea
f_S(E_{bb})=f_{SE}(E_{bb})-f_{DE}(E_{bb}) \label{eq:energydistaftermixed}
\eea
from which we will ultimately extract the rest-frame energy value  $E_{bb}^*$ for each fixed $m_{bb}$.
\begin{figure}[t]
\centering
\includegraphics[width=7.05cm]{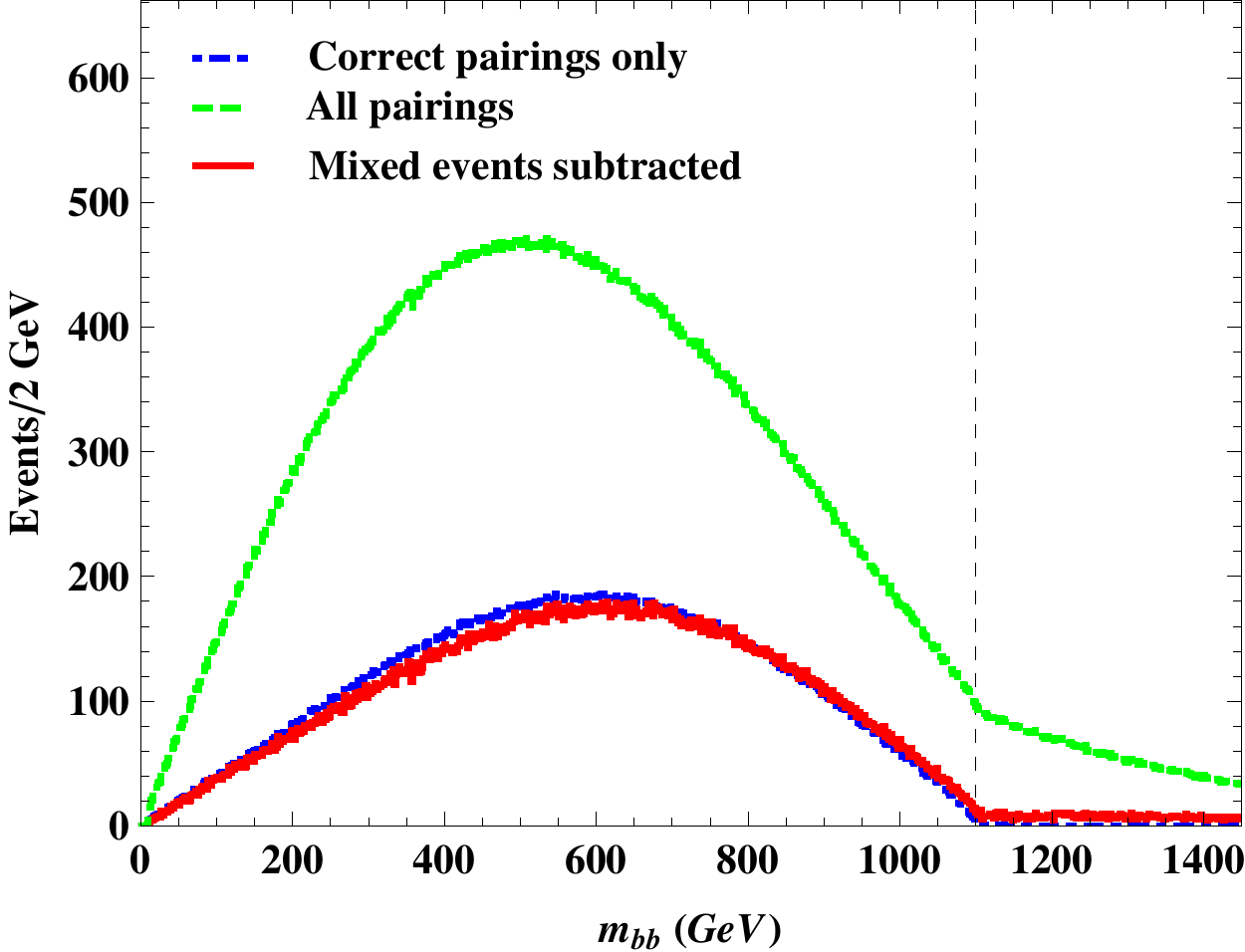}\hspace{0.5cm}
\includegraphics[width=7cm]{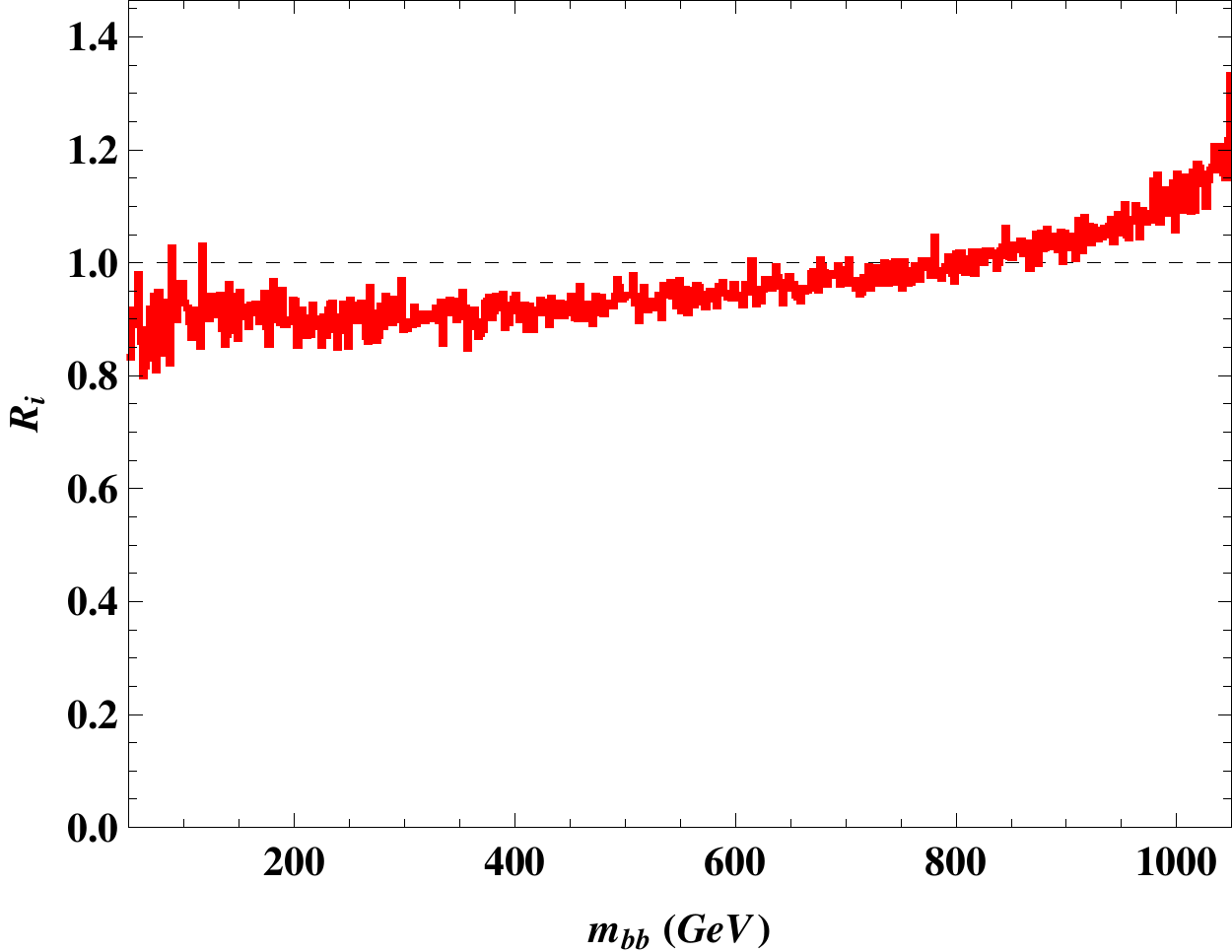}
\caption{\label{fig:puresig} The left panel shows the di-$b$-jet invariant mass distributions that are normalized to an integrated luminosity of $3\; ab^{-1}$. The right panel shows the $R_i$ distribution over $m_{bb}$. }
\end{figure}

Similarly to the subtraction for the $E_{bb}$ spectrum, we  obtain an estimate of the overall invariant mass distribution for the correct pairings using the mixed event technique, which is 
\bea
F_S(m_{bb})=F_{SE}(m_{bb})-F_{DE}(m_{bb})\,.
\eea 
Note that this distribution is not used to extract the masses, but serves as an example of the effectiveness of the mixed event subtraction scheme. To quantify the fidelity of the distribution obtained from the mixed event subtraction scheme, we define a ratio 
\bea
R_i\equiv \frac{N_{i,S}}{N_{i,C}}\,,\label{eq:chi2measure}
\eea
where $N_{i,S}$ is the $i$-th bin-count of the ``subtracted'' distribution $F_S(m_{bb})$ and $N_{i,C}$ is the corresponding bin-count in the distribution obtained considering only the correct pairs. The latter is obtained by exploiting the fact that in simulated events all of the history of the particles is available. 
 The left panel in Figure~\ref{fig:puresig} compares the  invariant mass distribution from the correct pairs, shown as the blue dot-dashed histogram, and the distribution obtained by mixed event subtraction, shown as the red solid histogram.  To show how much the original invariant mass distribution is contaminated by the combinatorial background, the distribution before the subtraction procedure is also plotted as the green dashed histogram. Each bin count is normalized to an integrated luminosity of 3 ab$^{-1}$. We observe that the distribution obtained from the mixed event subtraction is very close to that obtained from the correct pairs.  A more quantitative comparison is also provided in the right panel of Figure~\ref{fig:puresig}, showing the bin-by-bin ratio $R_i$ for the distribution of the events against $m_{bb}$. All of the bin counts are quite close to their associated theoretical values. In fact, $R_i \sim 1$ in all the range of $m_{bb}$.
We do not show the ratio $R_i$ in the vicinity of both kinematic endpoints because the significantly smaller $N_{i,C}$ at the endpoints leads to unreliable $R_i$ values. 
Besides visualizing the effectiveness of the mixed event subtraction, this check enables us to confirm that for a given invariant mass slice, a similar amount of data remain available after the mixed event subtraction compared to what would have been available were we able to completely eliminate the combinatorial background.

\begin{figure}
\centering
\includegraphics[width=7.1cm]{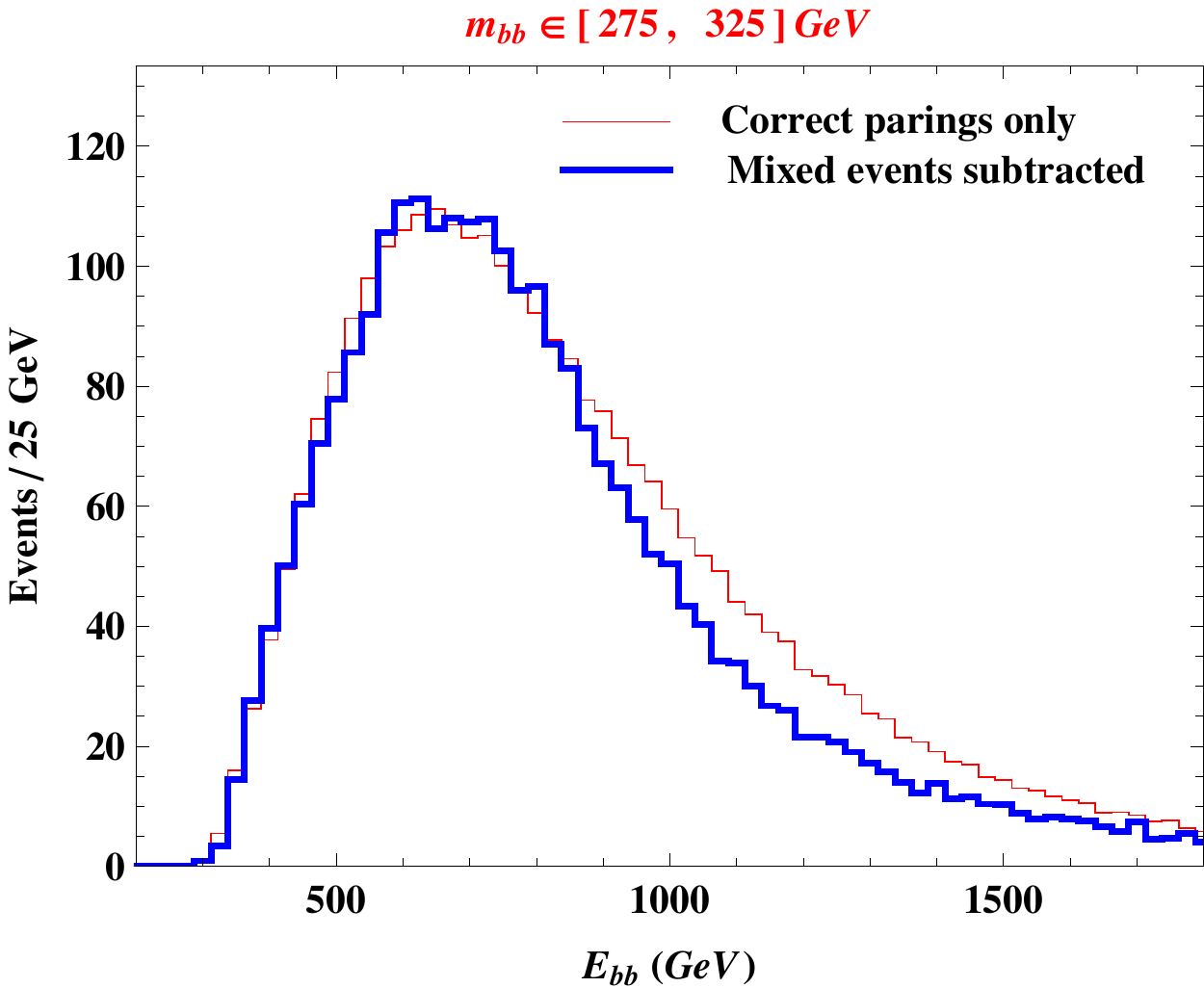}\hspace{0.5cm}
\includegraphics[width=7cm]{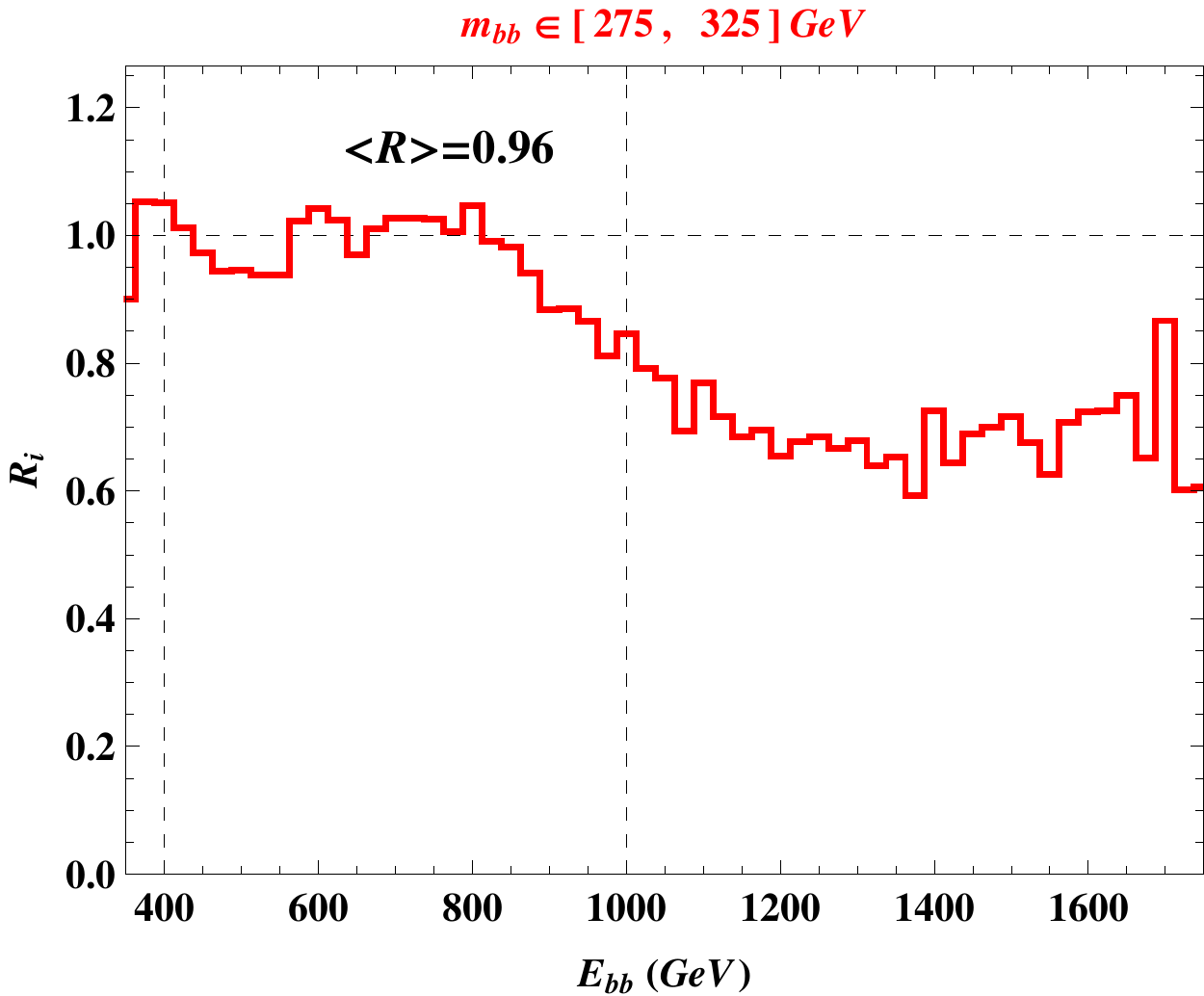}
\includegraphics[width=7cm]{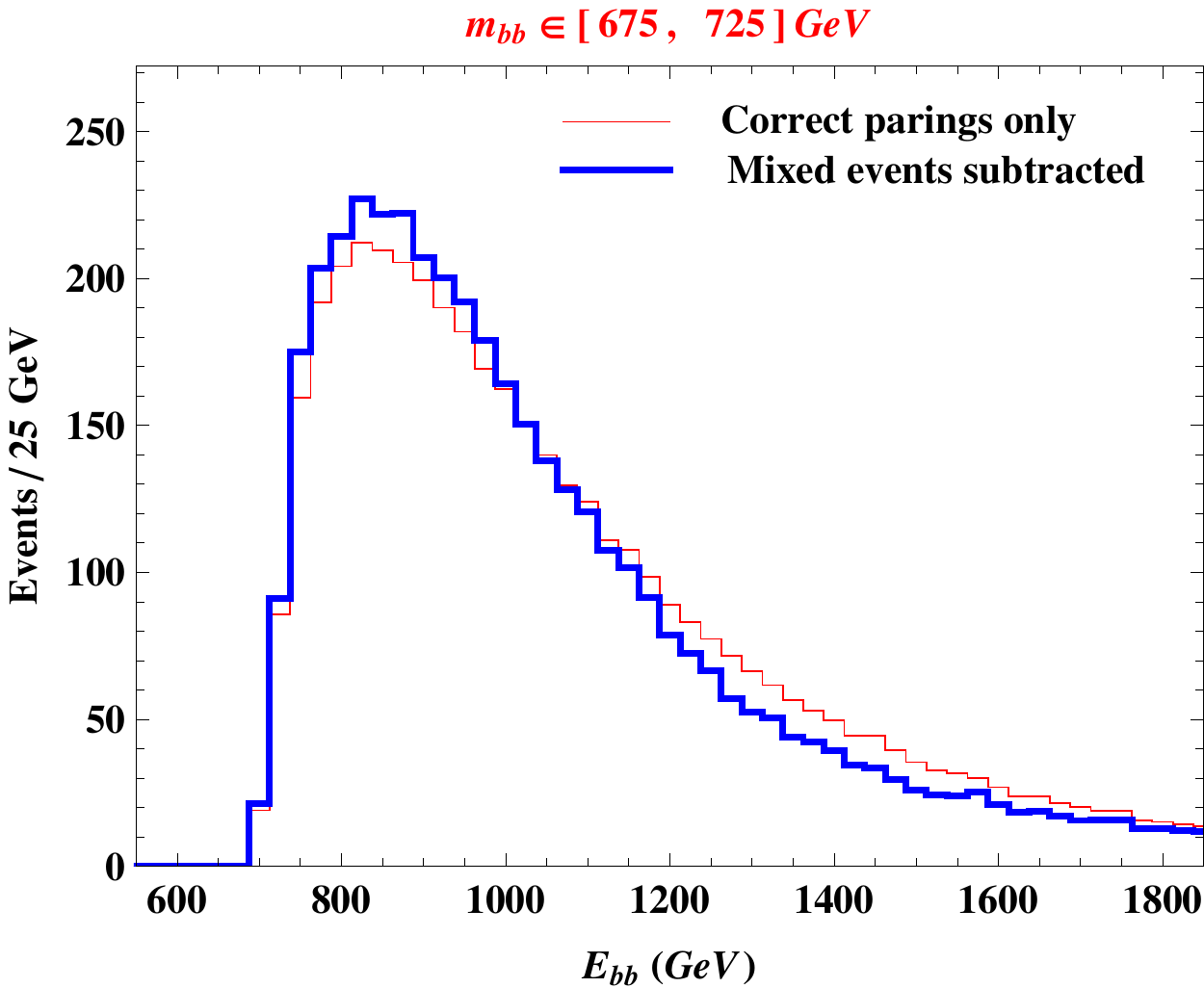}\hspace{0.5cm}
\includegraphics[width=7cm]{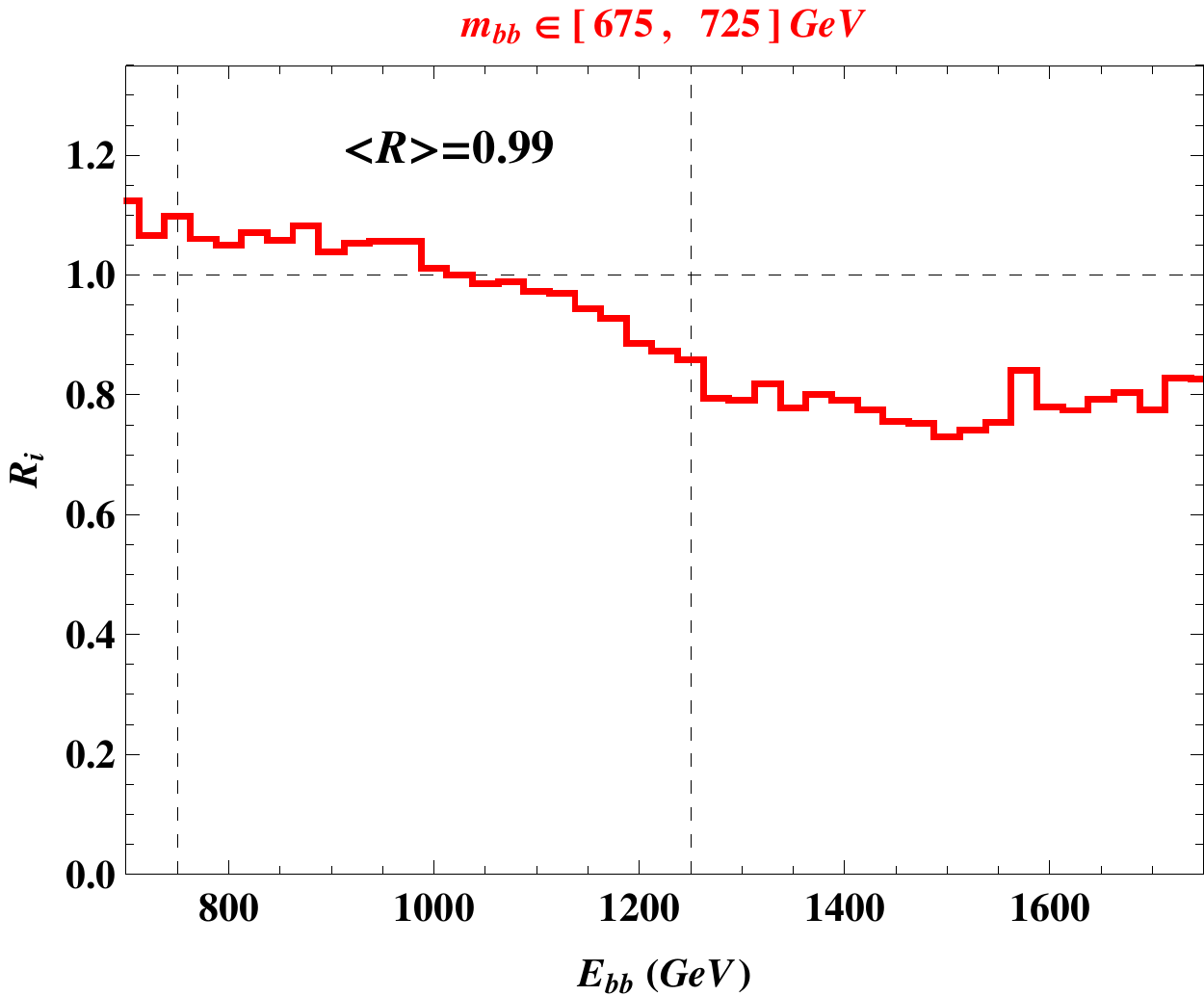}
\caption{\label{fig:energyMESS} The two plots in the left panel show the di-$b$-jet energy distributions for 300 GeV (top) and 700 GeV (bottom) nominal mass slices. They are normalized to an integrated luminosity of 3/ab. The color codes in the plots in the left panel are the same as those in Figure~\ref{fig:puresig}. The two plots in the right panel show the respective $R$ distributions over $E_{bb}$. For computing $\langle R\rangle$, only the data within the two black vertical dashed lines is taken into account.}
\end{figure}

In Figure~\ref{fig:energyMESS} we compare the energy distribution from the correct pairs and that obtained from the mixed event subtraction. The two distributions are shown for a mass slice $275\gev \leq m_{bb}\leq 325\gev$ in the upper panel of the figure and for another mass slice $675 \gev \leq m_{bb}\leq 725\gev$ in the bottom panel. The ratio of the correct pairs and the subtracted histograms is provided for each choice of $m_{bb}$ as well.
Since the fit to extract $E^{*}_{bb}$ from the energy spectra will be performed using only the data around the peak, we show the bin-by-bin ratio only for the energy range that corresponds to the full width at half maximum (FWHM), which is indicated by black dashed lines in each plot. In this case, we see that the energy spectrum  processed with the mixed event subtraction (blue histogram) is also quite close to the associated theory expectation (red histogram). To be more quantitative, we compute the average of $R_i$ in the FWHM range. This average is denoted as $\langle R\rangle$ and is close to 1, which suggests that the mixed event subtraction scheme works quite well, i.e., the shape of the energy spectrum is reasonably preserved. From this exploratory analysis, we expect that the extraction of $E_{bb}^*$ from the energy distribution obtained by the mixed event subtraction is unlikely to have major bias due to the subtraction. 

\paragraph{ii) Background and ``signal-background interference'':} Once the SM backgrounds come into play, there is a non-trivial complication that is introduced by the event mixing.
Since we are not aware {\it a priori} whether a given event is from the signal or the background, it is not possible to perform the event mixing using only the signal events. 
Therefore, the distribution returned by the whole operation of the mixed event subtraction scheme contains ``signal-background interference'', i.e., picking one particle from a signal event and the other from a background event. In principle, the overall kinematic characteristics of the background events differ from those of the signal events,  and 
therefore these interference pairs will make the overall distribution deviate from that of the pure signal or pure background combinatorially-generated background.\footnote{For the dominant background in our case (i.e., $Z+4b$), the pure background combinatorial distribution is somewhat tricky; the distinction between correct and incorrect pairings is meaningless because the associated event topology is ill-defined. However, in the interest of generality, we imagine that our background can also give rise to fictitious correct and wrong combinations.} As a consequence, a naively subtracted distribution would be distorted with respect to the distribution of a pure signal sample. 

To understand the quantitative impact of the inclusion of physical backgrounds, we first need to assess the hierarchy of the effects that arise from the simple addition of these backgrounds and from the event mixing. We focus on the situations where $n_{ev}$ events have been collected after the application of selection cuts, e.g., eqs.~(\ref{eq:cutforb})-(\ref{eq:delphicut}). These events come both from the signal process and from background processes. In general, we have $n_{s}$ signal events and $n_{b}=n_{ev}-n_{s}$ background events.\footnote{Since different types of backgrounds, in principle, will form different distributions, we here assume only a single type of background to avoid any potential complication.} However, in a situation in which a mass measurement is attempted, we expect that the signal will dominate the backgrounds, $n_{s}\gg n_{b}$.  Under this assumption, we can quantify how likely the event mixing procedure is to form pairs where both particles come from the signal process, both particles come from the background processes, or one particle comes from signal and the other from background.
Clearly, the probabilities to select a particle from a signal or a background event in the sample are 
\bea
p_{s}=\frac{n_{s}}{n_{s}+n_{b}}\simeq 1 - \frac{n_{b}}{n_{s}}\,,\quad\quad p_{b}=\frac{n_{b}}{n_{s}+n_{b}}\simeq \frac{n_{b}}{n_{s}},
\eea 
for a signal and a background event, respectively. Therefore, most of the pairs formed in the event mixing procedure are made with two particles from the signal process. Pairs made of two particles from the background are much less abundant, and in fact arise only in a small fraction, ${n_{b}^{2}}/{n_{s}^{2}}$, of the cases. Strikingly,  pairs made with one particle coming from the background and one particle from the signal are much more abundant than pure background pairs, as their probability is $2\times{n_{b}}/{n_{s}}$. 

The effect of the pairs involving both backgrounds and signal is not predictable unless one specifies the signal. However, some general features of this ``interference'' contribution to the event mixing estimate of the distribution for the correct signal pairings
can be easily guessed. First, the ``interference'' distribution tends to produce an underestimation of the bin-counts in the estimate eq.~(\ref{eq:eventmixingsubtraction}) of the distribution consisting of pairs of $b$-jets originating from the same decay. The reason is that in the inclusive distribution from all pairs in the same event, the first term in eq.~(\ref{eq:eventmixingsubtraction}), there are contributions stemming from pairs of events coming both from signal or both from background, but there is no way to construct a hybrid of the two: 
\bea
\frac{d\sigma}{dE_{bb}} (\textrm{all-pairs}) =  \frac{d\sigma_{S}}{dE_{bb}} +  \frac{d\sigma_{B}}{dE_{bb}},
\eea 
where by $\sigma_{S}$ and $\sigma_{B}$ we mean the signal and background contributions, respectively. On the other hand, in mixed events, we have 
\bea
\frac{d\sigma}{dE_{bb}} (\textrm{different-events pairs}) =  \frac{d\sigma_{SS}}{dE_{bb}} + \frac{d\sigma_{SB}}{dE_{bb}} + \frac{d\sigma_{BB}}{dE_{bb}}.
\eea
As suggested by Figure~\ref{fig:energyMESS}, the contribution from pairs where both events come from the signal,  $\frac{d\sigma_{SS}}{dE_{bb}}$, does a good job of estimating the effect of the pairs of $b$-jets not coming from the the same decay in the signal. Similarly, the contribution from pairs of events where both events come from the background, $\frac{d\sigma_{BB}}{dE_{bb}}$, is a good estimate of the combinatorial background generated from the background itself. Therefore, the contribution $\frac{d\sigma_{SB}}{dE_{bb}}$ is the piece that typically ruins the result because it gives rise to an excessive subtraction in eq.~(\ref{eq:eventmixingsubtraction}). Obviously, this phenomenon cannot be avoided since we are unable to distinguish signal and background events with absolute certainty. The presence of this type of ``interference'' background is inherent to the event mixing technique and dealing with it requires special care. A more quantitative argument about the interference is available in App.~\ref{sec:AppendixA} for more interested readers. 

\begin{figure}[t]
\centering
\includegraphics[width=7cm]{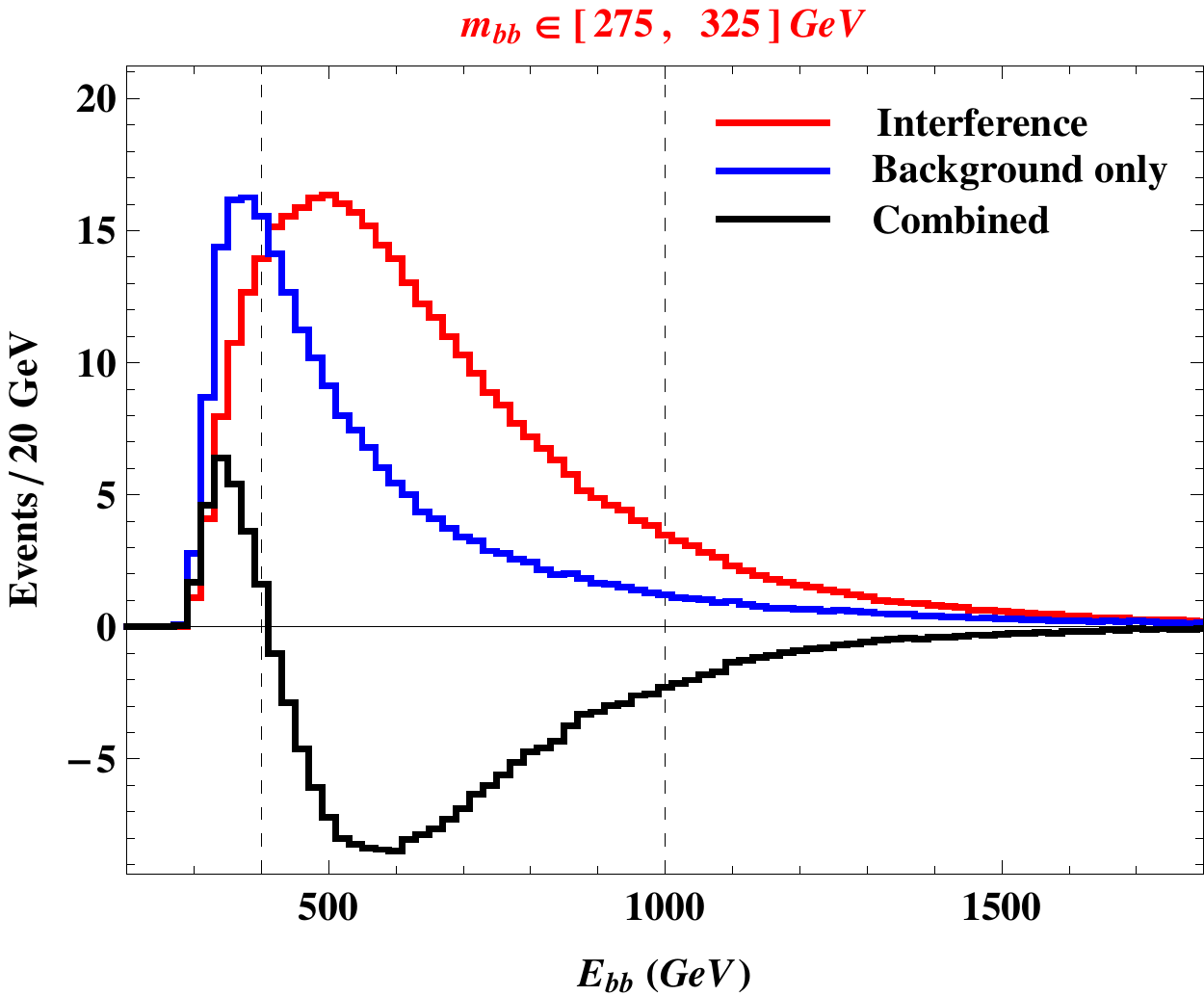}
\includegraphics[width=7cm]{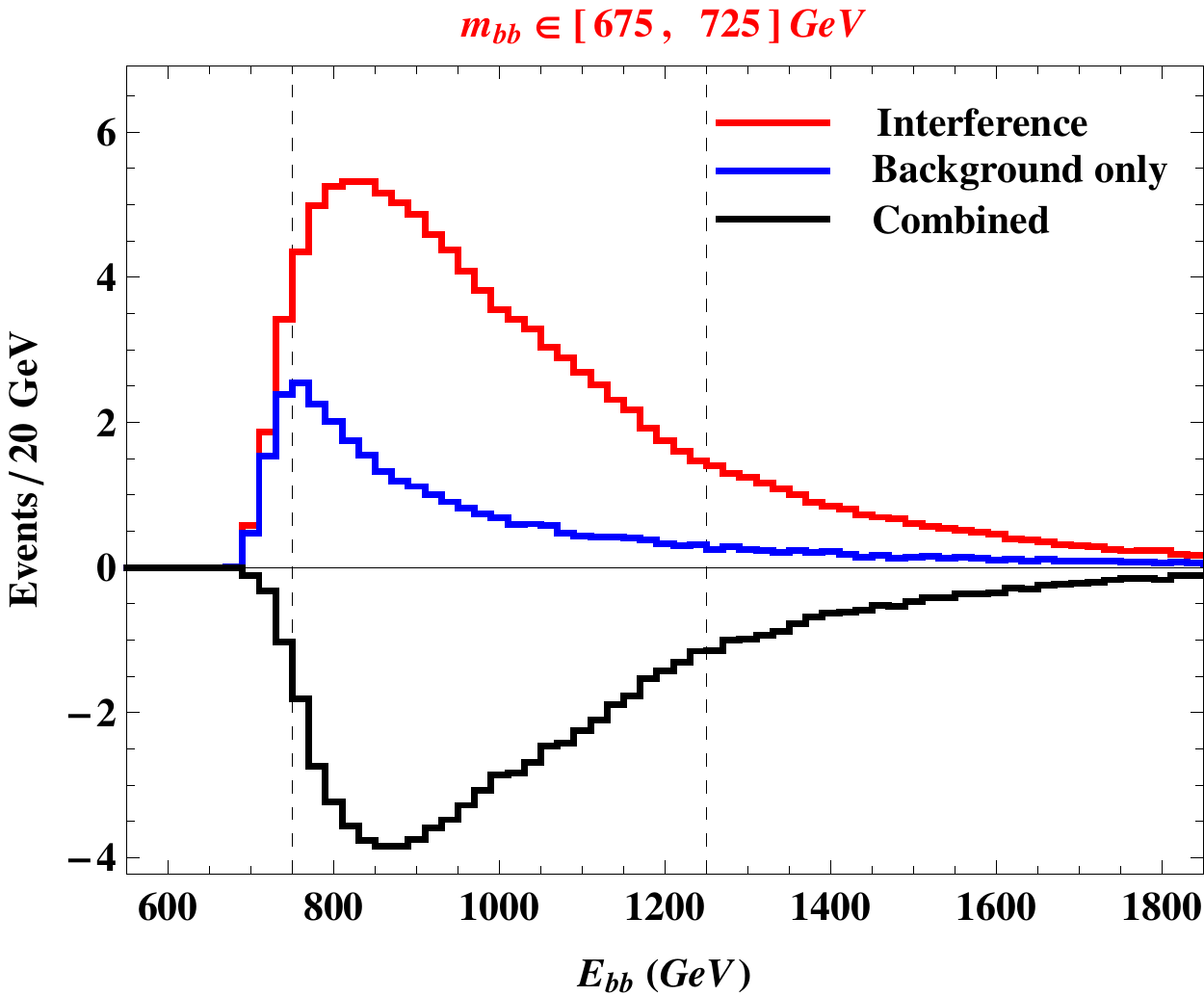}
\includegraphics[width=7cm]{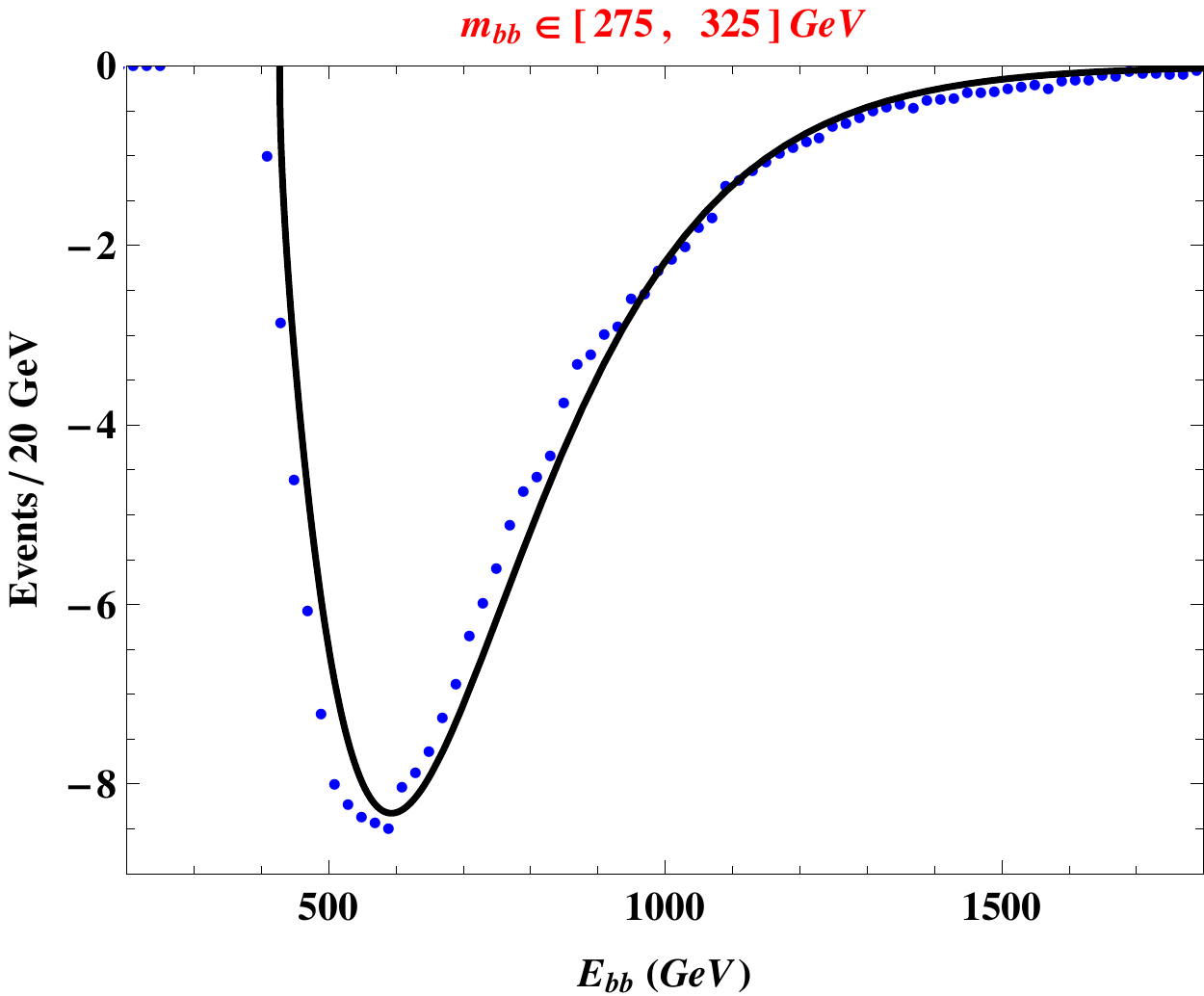}
\includegraphics[width=7cm]{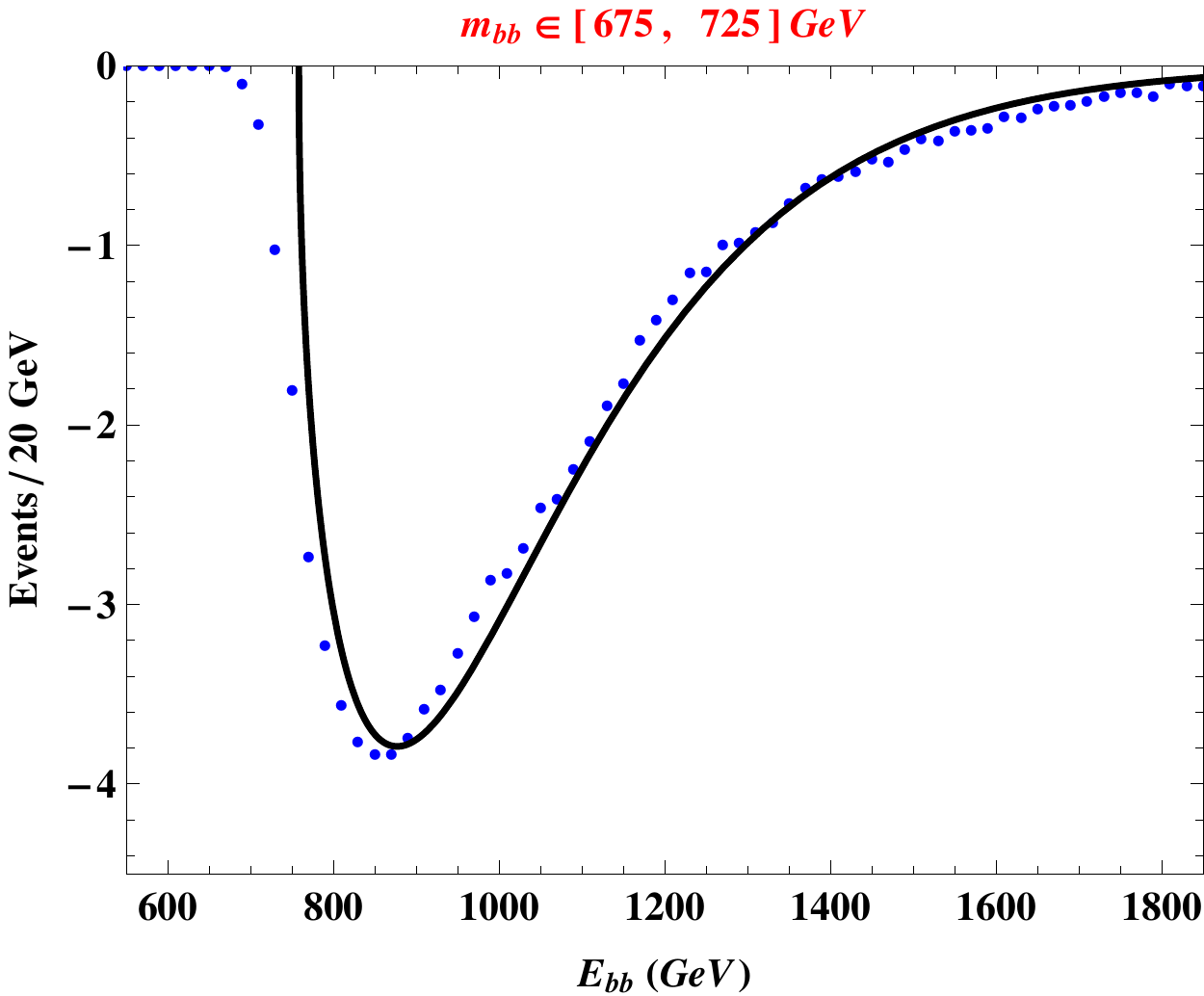}
\caption{\label{fig:interference} The di-$b$-jet energy distributions of the true background and the interference for 300 GeV (left panel) and 700 GeV (right panel) nominal mass slices. They are normalized to an integrated luminosity of 3 ab$^{-1}$. The black distribution is obtained by subtracting the blue one by the red one. The vertical black dashed lines denote the associated fitting range for each slice. The bottom panel shows the performance of the proposed fitting template for the effective backgrounds. }
\end{figure}

Given its peculiar origin, it may be desirable in some cases to remove the ``signal-background interference'' contribution. In order to do so, one must discuss the shape of this distribution, which in general depends on the signal and therefore is {\it a priori} unknown. However, some general features of the ``signal-background interference'' distribution can be predicted using the following argument. The distribution that arises from pairs made of one background and one signal particle feels in part the kinematics of the signal events and in part that of the background events. The background is typically expected to have softer particles than the signal, and therefore, the ``interference'' $b$-jet pair energy distribution is expected to be skewed towards energies that are somewhat larger than the characteristic values of the distribution for pairs made out of two background particles. For our example process, we display the distribution from pairs of signal particles in Figure~\ref{fig:energyMESS}, while the distribution from pairs of background particles and the interference distribution are shown in Figure~\ref{fig:interference}. Comparison of these distributions confirms our intuition from the argument above.

From Figure~\ref{fig:interference} we see that the interference effect dominates the pure background effect essentially everywhere in the distribution. As discussed above, this is due to the fact that we envision a situation where signal events are much more abundant than background events. The shape of the interference contribution is also quite different from that of the pure background contribution. 

When discussing results in a later section, we will study the effect of the removal of background contributions to our results. With this goal in mind, we study what functional form describes the {\it total} effect of backgrounds; that is, the distribution from the pure background pairs plus that from the ``signal-background interference'' pairs. The combination is shown in Figure~\ref{fig:interference}; in the lower panel, we show a possible fit of this distribution. Due to the importance of the signal in determining the shape of the ``signal-background interference'' distribution, we decided to model the {\it total} effect of backgrounds with a function of the family eq.~(\ref{eq:massiveTemplate}). The fit result in Figure~\ref{fig:interference} is rather good, but we do not attach any special significance to this finding. In fact, a better description for this background may exist and might be preferred. More generally, we stress that the ``signal-background interference'' distribution is not universal, and our choice could be unreliable for other signals. In our application to the gluino decay process, the fairly good description provided by eq.~(\ref{eq:massiveTemplate}) and shown in Figure~\ref{fig:interference} is satisfactory for our current purposes.

\section{Mass measurement results and discussion\label{sec:results}}
\begin{figure}[t]
\centering
\includegraphics[width=7cm]{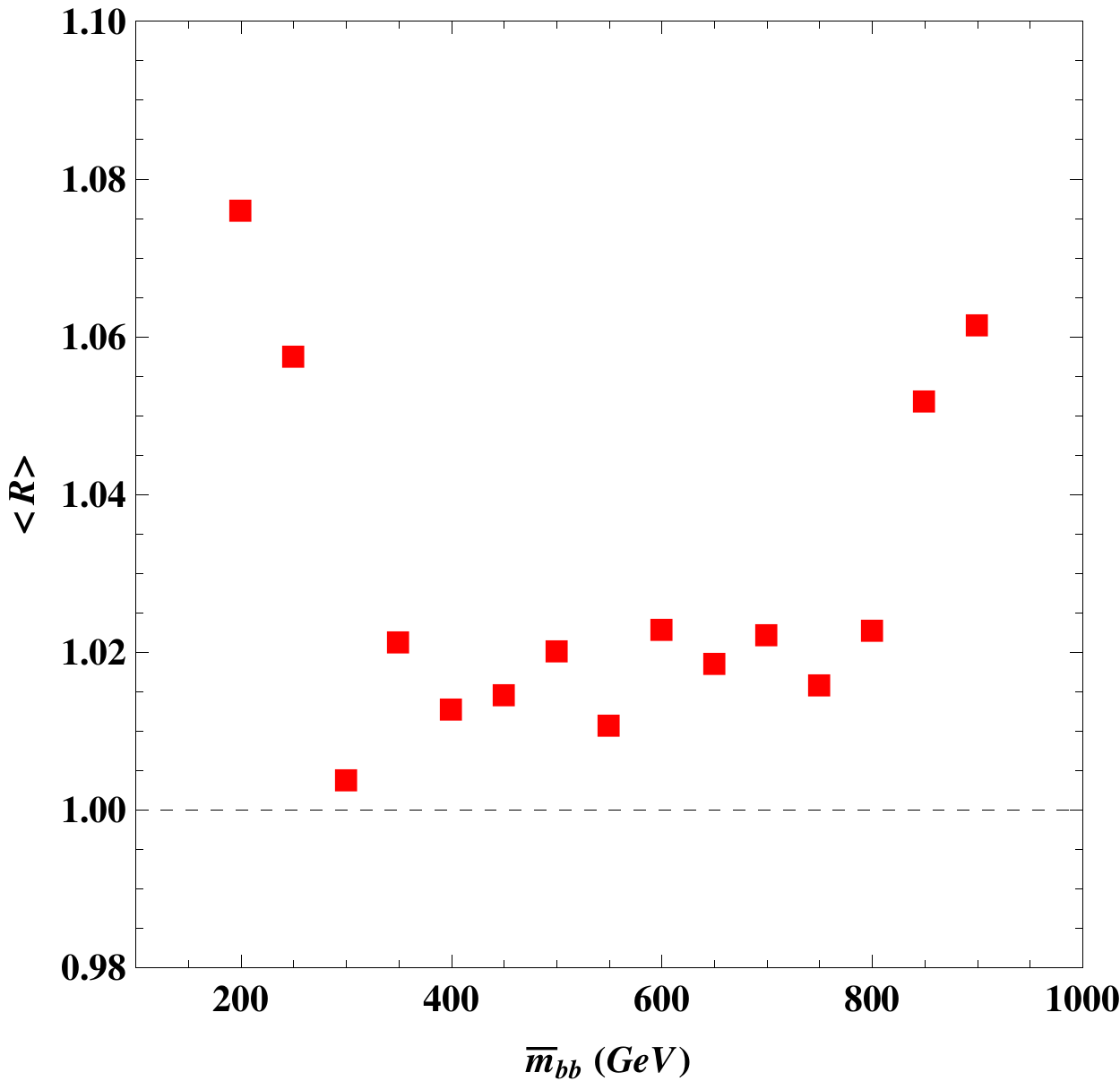}\hspace{0.5cm}
\caption{\label{fig:Rratio} Average in the FWHM range of the bin-by-bin ratio of the  energy distributions from the event mixing and from just the correct pairs of $b$-jets.  %% 
$\langle R\rangle=1$ implies a good match.}
\end{figure}
In this section, we demonstrate the application of the proposed technique to the gluino decay. Results on the mass measurement from fitting the energy spectra for the compound system of two $b$-jets are presented in the following subsections along with the possible issues and limitations of our method. In the final subsection, we discuss possible improvements of the mass measurement with the aid of the di-jet invariant mass endpoint.
 
\subsection{Measurement of gluino and neutralino masses\label{sec:baselinemethod}}

Following the strategy outlined in the previous sections, we present results for the determination of the masses of the gluino and the neutralino from the $b$-jet energies. The energy spectra that we study are obtained from simulated event samples generated as described in Sec.~\ref{sec:signaldetails} for both signal and dominant background processes at the 14 TeV LHC. We also recall that the relevant channel is characterized by a large missing transverse momentum and four bottom-tagged jets which are selected as per eqs.~(\ref{eq:cutforb})-(\ref{eq:delphicut}). Since the primary interest of this paper is to study the theoretical aspects of energy peaks in a multi-body decay, rather than data analysis under realistic statistics, we take a sufficiently large number of events to minimize potential statistical fluctuation within the data sample, which is then normalized to an integrated luminosity of 3 ab$^{-1}$. 

\begin{figure}[t]
\centering
\includegraphics[width=7cm]{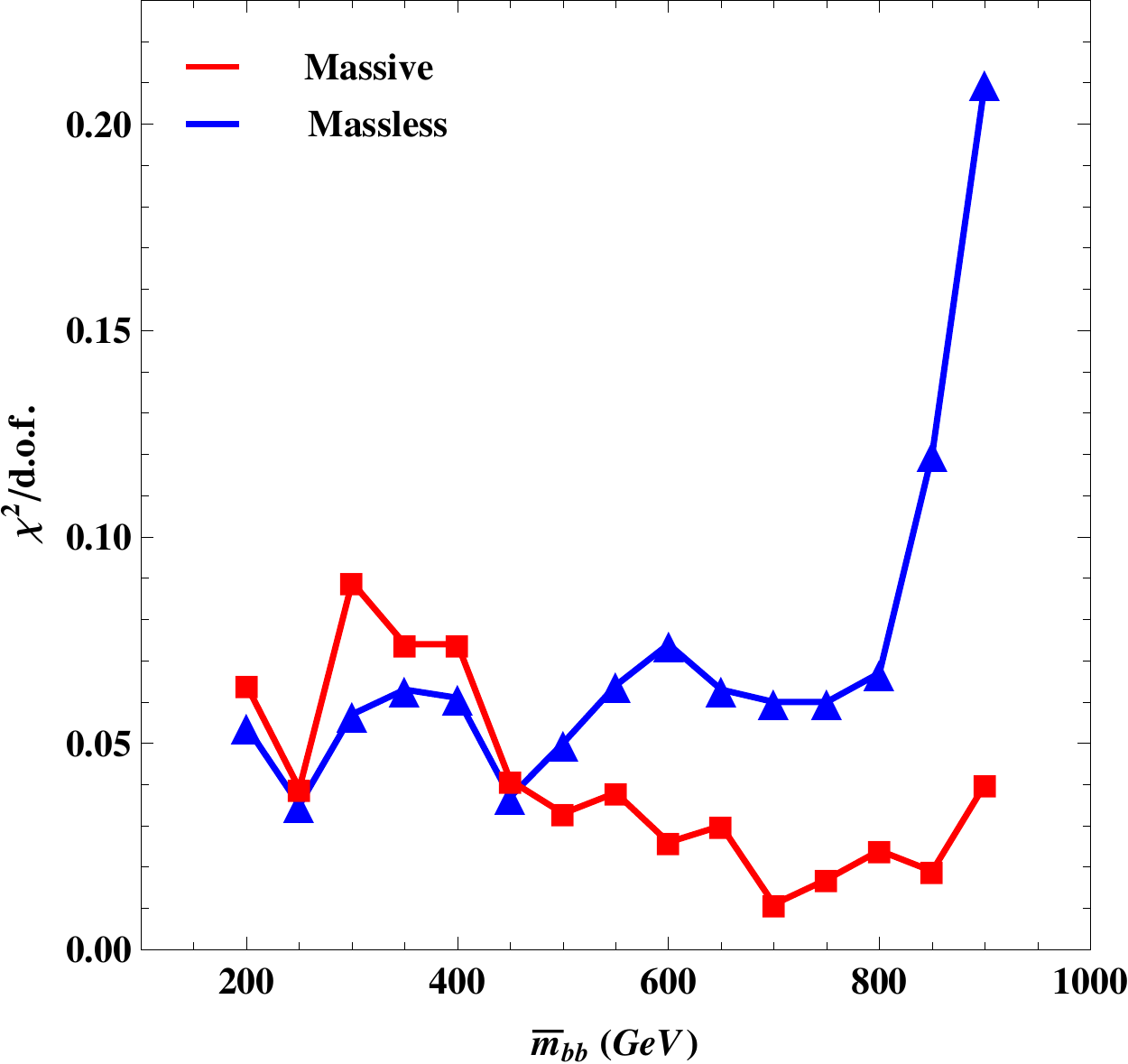}
\caption{\label{fig:chiratio} Comparison of the goodness of the fit between massive template eq.~(\ref{eq:massiveTemplate}) (blue) and the massless template eq.~(\ref{eq:masslessTemplate}) (red) as fitter for the energy spectrum of the $b$-jet pair system for various values of the  invariant mass of $b$-jet system $\bar{m}_{bb}$.  }
\end{figure}

Note that we study the distribution of the sum of the energy of $b$-jet pairs (say, $b_1$ and $b_2$), i.e., $E_{bb}=E_{b_1}+E_{b_2}$, for which 
the associated invariant mass values belong to a narrow range such that
\bea
m_{bb} \in [\bar{m}_{bb}-\Delta m_{bb}/2,\;\bar{m}_{bb}+\Delta m_{bb}/2]\,,
\eea
where $\Delta m_{bb}$ denotes the width of the mass window. We henceforth identify every individual mass window by its central value $\bar{m}_{bb}$. Due to the indistinguishable nature of the final state particles, we form the $E_{bb}$ distribution from all possible pairs of $b$-jets in an event, and subsequently apply the mixed event subtraction technique described in Sec.~\ref{sec:MESS} to eliminate the contamination from the pairs of $b$-jets {\it not} coming from the same gluino.
In Figure~\ref{fig:Rratio}, we present the average of the bin-by-bin ratio of the distributions from only correct pairs and from the mixed event subtraction technique at various values of $\bar{m}_{bb}$. We remark that for each point in the figure, the average is limited to the energy range defined by the full width at half maximum (FWHM) of the distribution. The figure suggests that the average deviation between the two distributions is, at most, about 8\%, meaning that the mixed event subtraction scheme reproduces the original distribution fairly well. Although we do not show it here, we also remark that for a given $\bar{m}_{bb}$ the standard deviation in the bin-by-bin ratio is small enough that the distribution from the mixed event subtraction consistently tracks the corresponding distribution from the correct pairs.

For each $\bar{m}_{bb}$, the rest-frame energy of the $b$-jet pair system (i.e., $E_{bb}^*$) is extracted from the energy distribution by fitting the data to a template function. We have two possible template functions, given in eqs.~(\ref{eq:masslessTemplate}) and~(\ref{eq:massiveTemplate}), and we use both of them to see which one better suits the data and to see if there are significant differences between the rest-frame energy and $E^*$ value found by the two functions. Since eq.~(\ref{eq:masslessTemplate}) is suitable only for the case where the relevant $\bar{m}_{bb}$ is effectively negligible, we expect an increasing discrepancy between the results obtained by eqs.~(\ref{eq:masslessTemplate}) and~(\ref{eq:massiveTemplate}) as $\bar{m}_{bb}$ grows.  
The reduced $\chi^{2}$  values for each fit to the energy spectrum are shown in Figure~\ref{fig:chiratio}. This $\chi^{2}$ is a measure of how well the template function describes the data globally, but we do not necessarily attach any statistical meaning to it. We instead use it as a measure for the distance between the two template functions. Looking at the figure, we observe that for all $\bar{m}_{bb}$, the massive template describes the data as well as or better than the massless template. As expected, the performance of the massless template becomes progressively worse as $\bar{m}_{bb}$ increases. 
The massless template also seems inferior to the massive template from another aspect as it typically returns an estimate of the rest-frame energy that is larger than the  expected value. On the other hand, the massive template does not introduce such a pronounced bias.

\begin{figure}[t!]
\centering
\includegraphics[width=7cm]{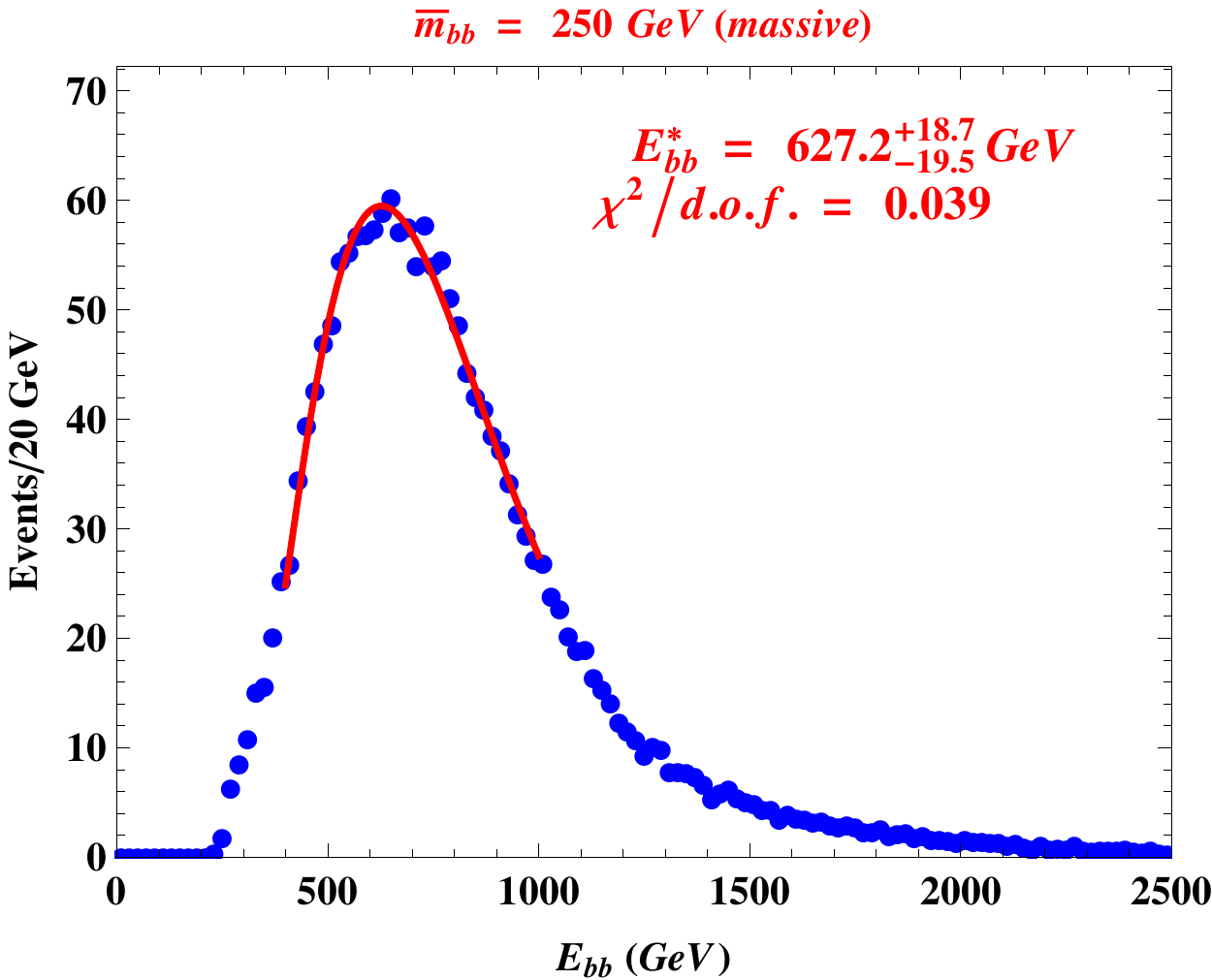} \hspace{0.5cm}
\includegraphics[width=7cm]{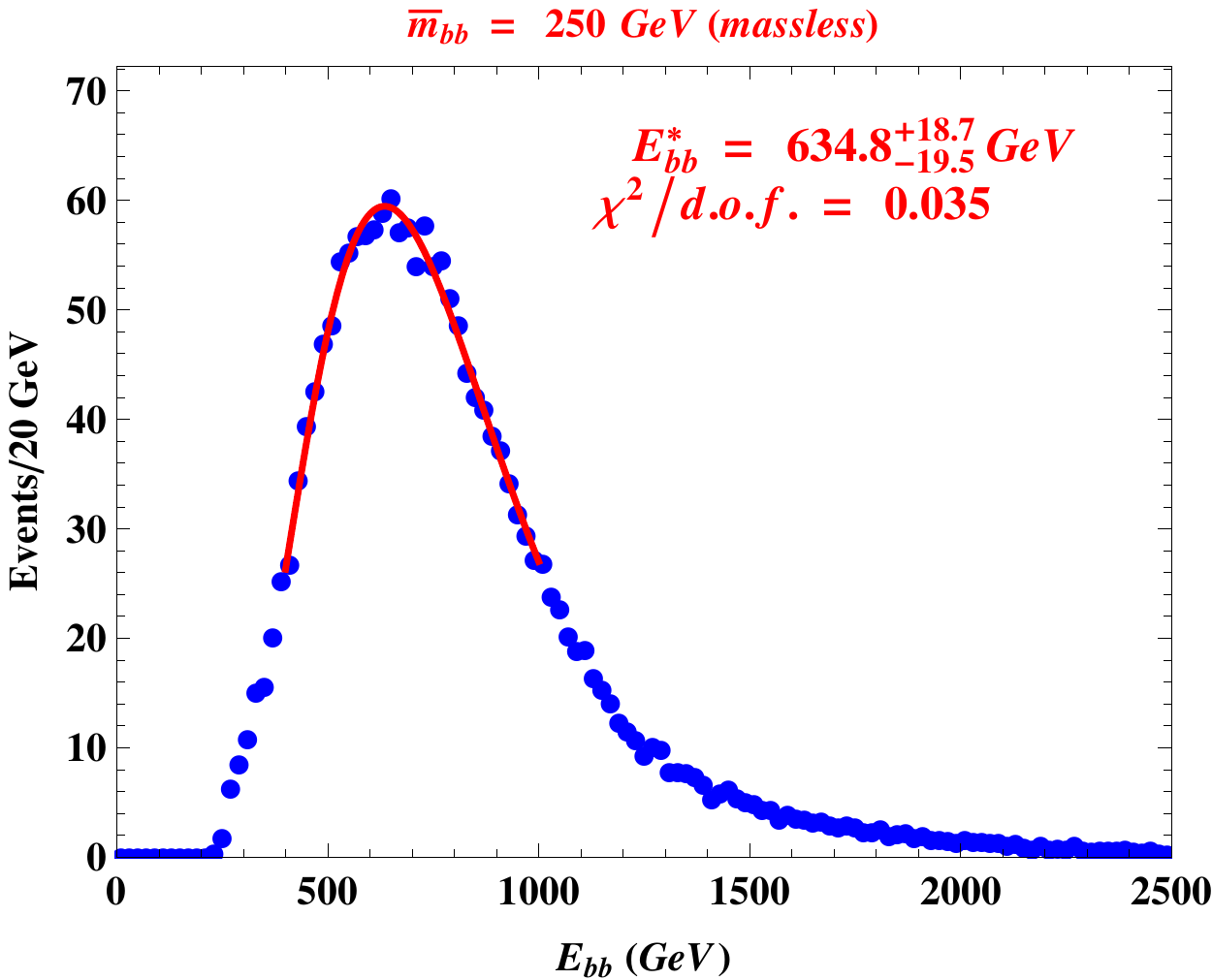}
\includegraphics[width=7cm]{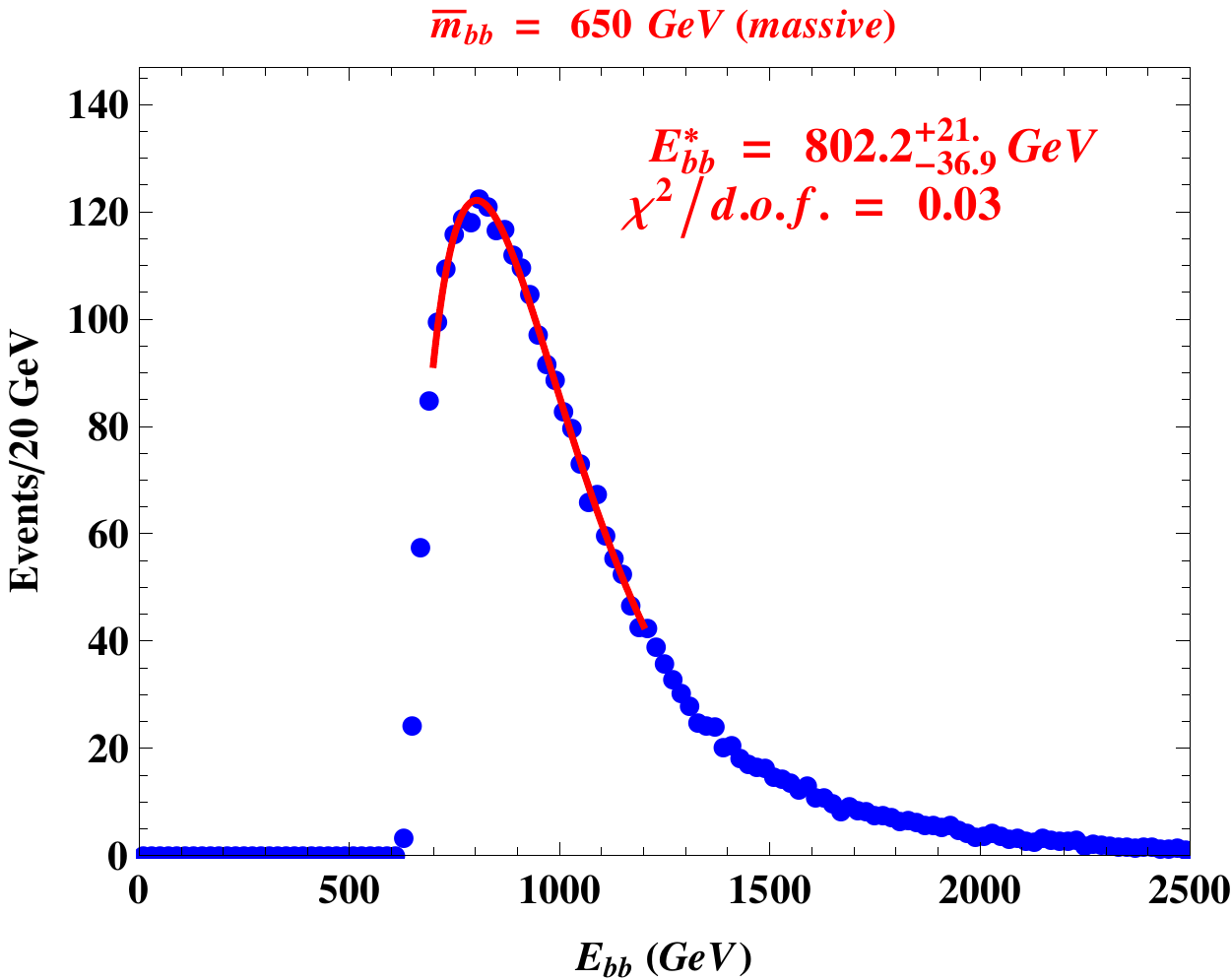} \hspace{0.5cm}
\includegraphics[width=7cm]{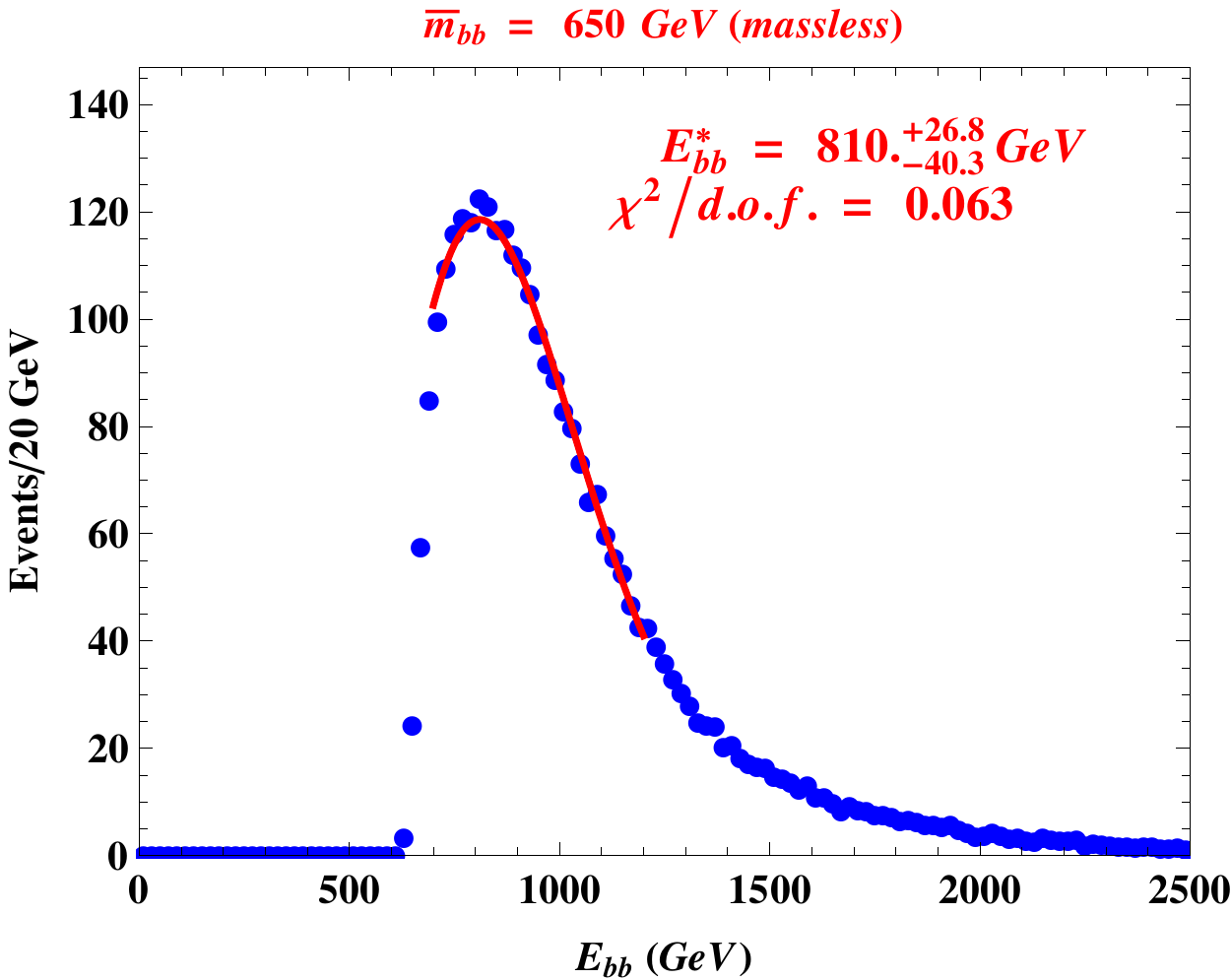}
\caption{\label{fig:samplefits} Sample fit results for extracting $E_{bb}^*$ using the massless template (right panels) and the massive template (left panels). The chosen invariant mass slices are $m_{bb}\in[225,\;275]$ GeV (top panels) and $m_{bb}\in[625,\;675]$ GeV (bottom panels). We report only  statistical errors.
Each fit range is chosen such that it roughly corresponds to the relevant FWHM.}
\end{figure}

To demonstrate the difference between fits made with the two templates, we show in Figure~\ref{fig:samplefits} sample fit results for two different nominal $\bar{m}_{bb}$ values, 250~GeV and 650~GeV. In the left panels, we provide results from applying the massive template, while in the right panels we provide results  from applying the massless template. 
The errors quoted for the extracted $E_{bb}^*$ were estimated at  95\% confidence interval (C.I.)  from  the variation of the $\chi^2$  of the fit. 
 For both of the $\bar{m}_{bb}$ values, we see that the massless template estimates $E_{bb}^*$ as being slightly larger than the estimate given by the massive template, and the discrepancy is larger as we go to larger $\bar{m}_{bb}$. Although the discrepancy is within the 95\% confidence interval, we feel that this is an important characteristic of the fit results. In fact, the massless template has a systematic tendency to return larger $E_{bb}^{*}$ in general, which implies the introduction of a possible bias to the mass measurement. The results of the fits for all of the values of $\bar{m}_{bb}$ are reported in Table~\ref{tab:fitresults}, from which we see that massless template consistently overshoots the estimate of $E_{bb}^{*}$ obtained from the massive template, and that it also overestimates the theory values of $E_{bb}^{*}$.  
 
\begin{table}[t]
\centering
\begin{tabular}{c|c||c c c}
$\bar{m}_{bb}$ & Fit range & Theory & Massive [$\chi^2$/d.o.f] & Massless [$\chi^2$/d.o.f] \\
\hline \hline 
 & & & \\
200 & [400, 1000] & 612.5 & $614.8_{-24.0}^{+21.4}$ [0.064] & $619.1_{-24.2}^{+21.4}$ [0.054] \\
250 & [400, 1000] & 621.9 & $627.2_{-19.5}^{+18.7}$ [0.039] & $634.8_{-19.5}^{+18.7}$ [0.035] \\
300 & [400, 1000] & 633.3 & $640.9_{-16.0}^{+15.5}$ [0.089] & $654.3_{-16.0}^{+15.4}$ [0.057] \\
350 & [440, 1000] & 646.9 & $659.2_{-16.2}^{+16.2}$ [0.074] & $673.1_{-16.2}^{+15.9}$ [0.063] \\
400 & [440, 1040] & 662.5 & $670.7_{-14.0}^{+14.3}$ [0.074] & $693.7_{-13.5}^{+13.6}$ [0.061] \\
450 & [500, 1040] & 680.2 & $694.8_{-15.5}^{+14.9}$ [0.041] & $715.9_{-15.6}^{+14.8}$ [0.037] \\
500 & [540, 1040] & 700.0 & $716.3_{-15.9}^{+14.8}$ [0.033] & $738.8_{-16.2}^{+14.8}$ [0.050] \\
550 & [600, 1100] & 721.9 & $742.1_{-21.1}^{+16.7}$ [0.038] & $760.0_{-23.8}^{+18.5}$ [0.064] \\
600 & [640, 1100] & 745.8 & $768.9_{-27.2}^{+17.6}$ [0.026] & $787.9_{-26.3}^{+19.3}$ [0.074] \\
650 & [700, 1200] & 771.9 & $802.2_{-36.9}^{+21.0}$ [0.030] & $810.0_{-40.3}^{+29.0}$ [0.063] \\
700 & [740, 1240] & 800.0 & $832.7_{-132.7}^{+23.4}$ [0.011] & $840.9_{-45.0}^{+21.4}$ [0.060] \\
750 & [800, 1300] & 830.2 & $871.4_{-121.5}^{+28.4}$ [0.017]& $865.3_{-68.7}^{+39.3}$ [0.060] \\
800 & [840, 1340] & 862.5 & $910.5_{-110.6}^{+28.3}$ [0.024] & $908.3_{-64.3}^{+37.8}$ [0.067] \\
850 & [880, 1340] & 896.9 & $952.8_{-102.9}^{+29.4}$ [0.019] & $961.1_{-60.7}^{+34.4}$ [0.12] \\
900 & [920, 1400] & 933.3 & $998.0_{-98.0}^{+29.7}$ [0.040] & $1015.2_{-52.6}^{+31.6}$ [0.21]\\
 & & & \\
\hline
\end{tabular}
\caption{\label{tab:fitresults} The fit results for fifteen invariant mass slices. For each fit, a mass slice of 50 GeV was chosen, for example, for $\bar{m}_{bb}=200$, $E_{bb}$ is selected such that the corresponding $m_{bb}$ is between 175 and 225 GeV. The bin size for all energy distributions is 20 GeV. The error estimation for each fit parameter is performed by 95\% confidence interval. All values but $\chi^2$ are given in GeV.}
\end{table}

%%%%%%%%%%%%%%%%%%%%%%%%%%%%%%%%%%%%
%%%%%%%%%%%%%%%%%%%%%%%%%%%%%%%%%%%% 
\begin{figure}[t]
\centering
\includegraphics[scale=1]{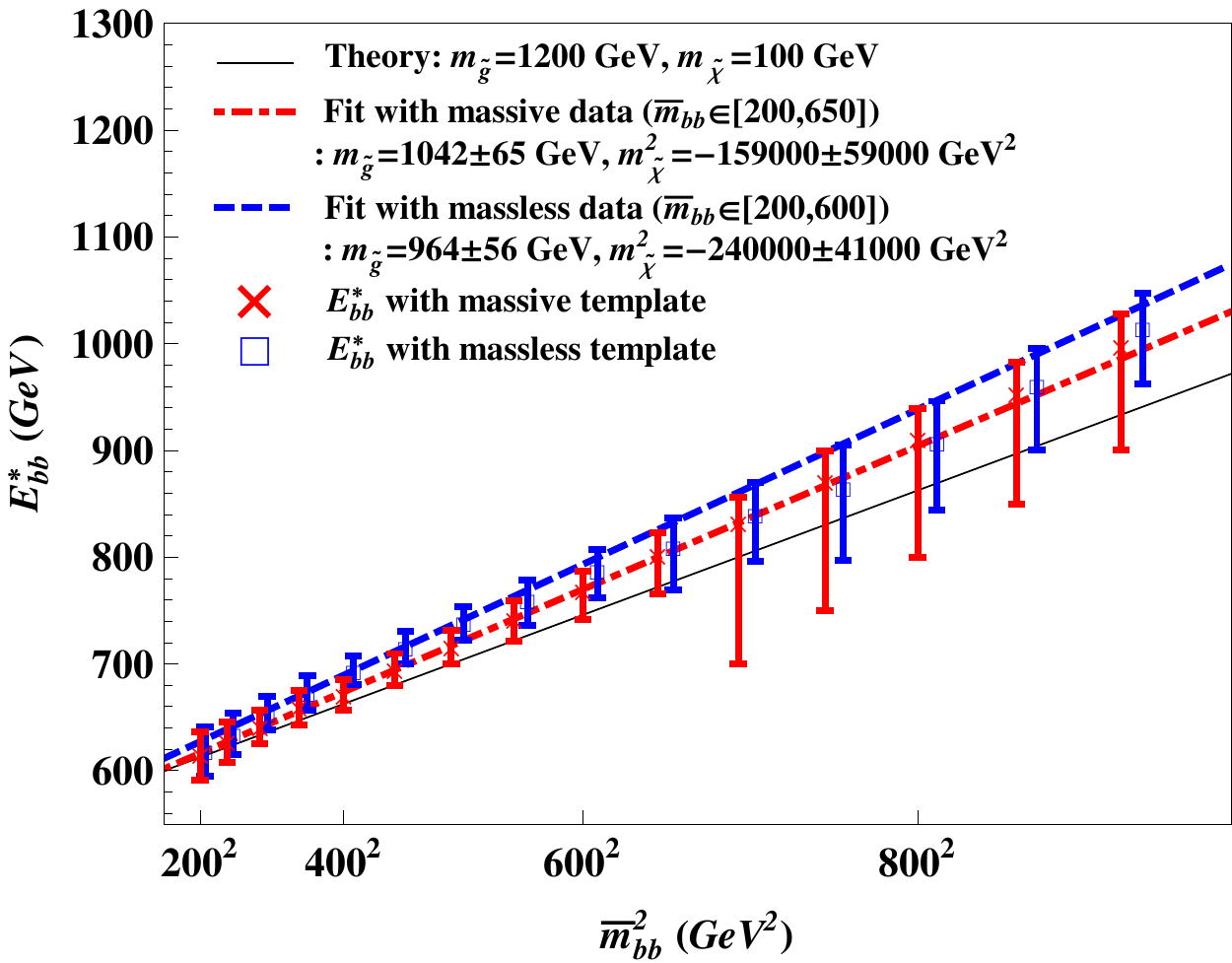}
\caption{\label{fig:MeasureData} The fit of the data points $(m_{bb}^2,\;E_{bb}^*)$ with eq.~(\ref{eq:lineformeasurement}). The theoretical expectation for the given mass spectrum is represented by the solid black line. The data points obtained by fitting the $E_{bb}$ distributions with massless and massive templates are marked by ``cross'' and ``square'' symbols, respectively. The mass measurement done with cross symbols is represented by the red dashed line. For the blue dot-dashed line, the measurement is done for the data points for which the massless template work reasonably well.}
\end{figure}
\begin{figure}[t]
\centering
\includegraphics[width=6.7cm]{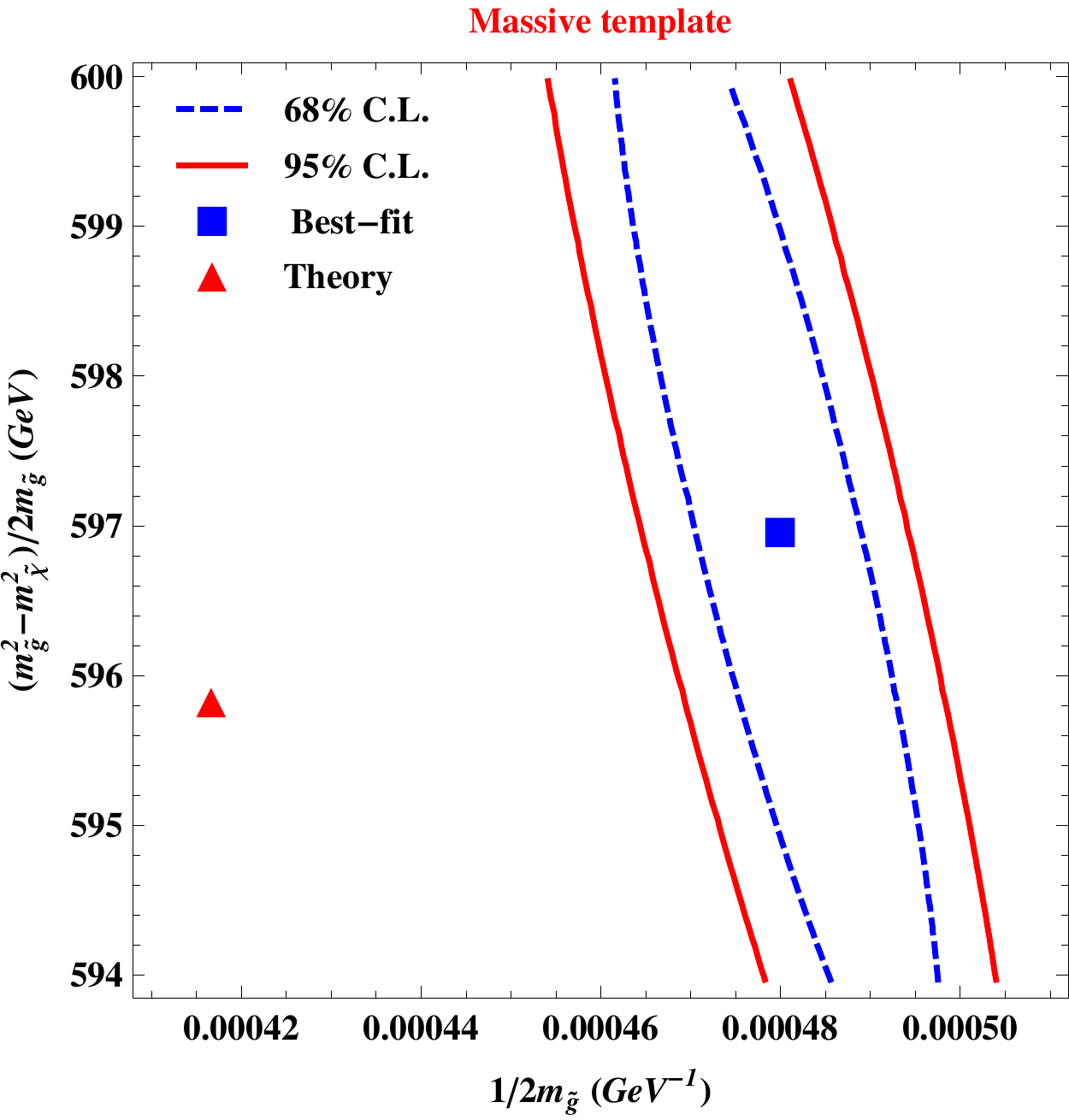}\hspace{0.5cm}
\includegraphics[width=7cm]{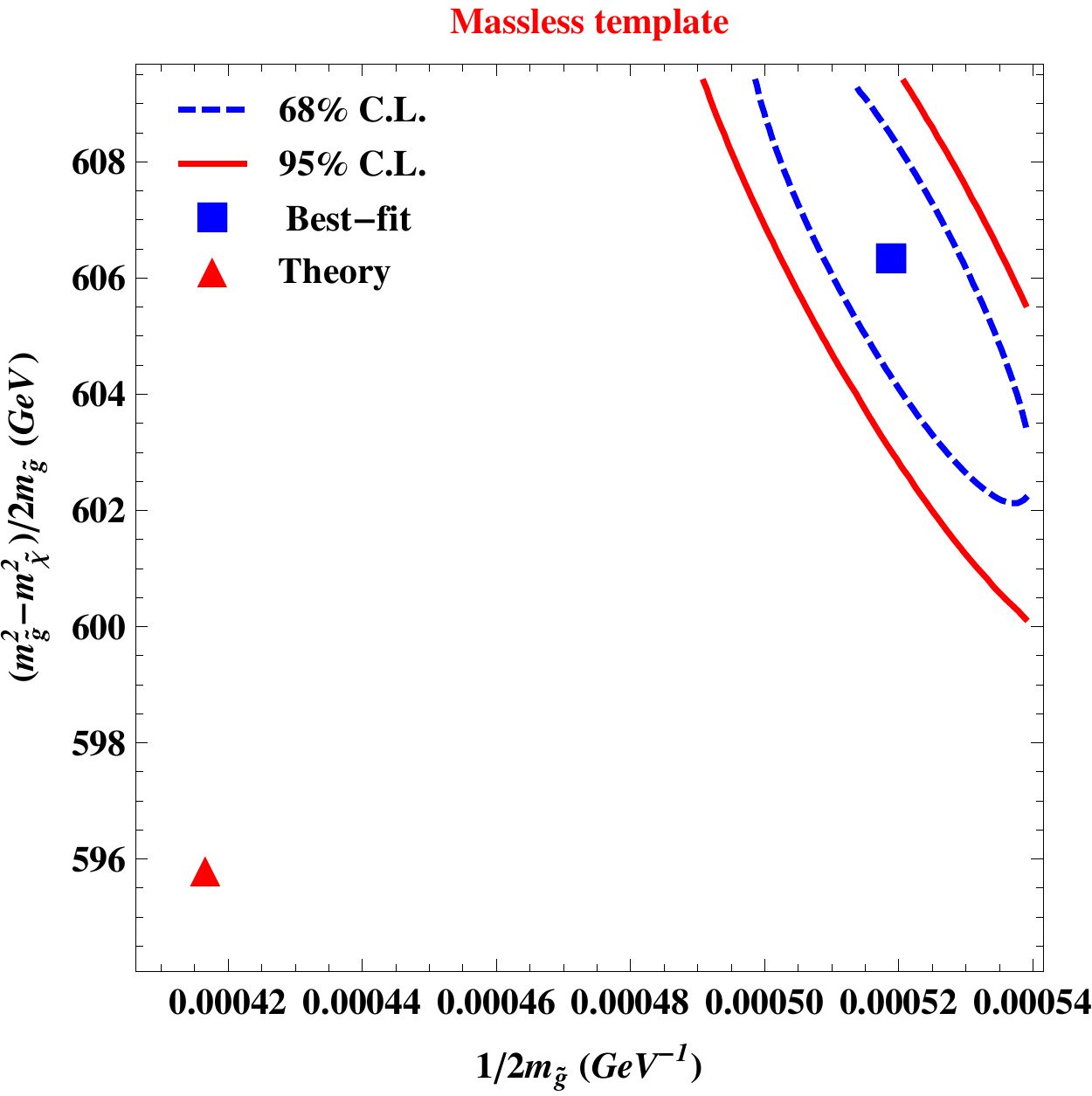}
\caption{\label{fig:contour} Contour plots in the plane of \texttt{s} ($=1/2m_{\tilde{g}}$) vs. \texttt{y} ($=(m_{\tilde{g}}^2-m_{\tilde{\chi}}^2)/2m_{\tilde{g}}$) around the best fit values for the fit results with massive (left panel) and massless (right panel) templates. }
\end{figure}

%%%%%%%%%%%%%%%%%%%%%%%%%%%%%%%%%%%%%%%
%%%%%%%%%%%%%%%%%%%%%%%%%%%%%%%%%%%%%%%
Finally, we take all the values of $E_{bb}^*$ obtained from the fitting the energy spectrum for each $\bar{m}_{bb}$ and fit them to the line given by eq.~(\ref{eq:massiveestarrest}), taking into account and displaying the associated errors from the fit procedure that we used to extract the values of $E_{bb}^*$. 
The expression in eq.~(\ref{eq:massiveestarrest}) can be adjusted to match our example process as follows: 
\beq
E_{bb}^{*} = \frac{m_{\tilde{g}}^{2}-m_{\chi}^{2}+\bar{m}_{bb}^{2}}{2m_{\tilde{g}}} \,.\label{eq:lineformeasurement}
\eeq
We perform the fit of eq.~(\ref{eq:lineformeasurement}) on both of the results from the massive and massless templates. For the fits using the massive template, we use only the results obtained for $\bar{m}_{bb}$ in the range from 200~GeV to 650~GeV, in which the errors from the fit of the energy spectra are quite small. Other choices of the $\bar{m}_{bb}$ range give similar mass measurements, but we simply make a conservative choice for the range of $\bar{m}_{bb}$ included so to avoid the values of $\bar{m}_{bb}$ where fewer events are expected. From the results extracted using the massless template, we choose to fit eq.~(\ref{eq:lineformeasurement}) only for $\bar{m}_{bb}$ in the range from 200~GeV to 600~GeV, where it is more reasonably accurate to treat the $b$-jet pair system as  massless. The fit parameters are the slope of eq.~(\ref{eq:lineformeasurement}) and its vertical intercept (that is $E_{bb}^{*}$ of $\bar{m}_{bb}=0$) and we denote them as  
\beq  \texttt{s} \equiv \frac{1}{2m_{\tilde{g}}}\quad\textrm{ and }\quad \texttt{y} \equiv \frac{m_{\tilde{g}}^2-m_{\tilde{\chi}}^2}{2m_{\tilde{g}}}\,.\label{eq:observablestomasses}\eeq For our spectrum, the theory values are
\beq
\texttt{s}=4.2\times 10^{-4}\gev^{-1},\quad \texttt{y}=595 \gev\,. \label{eq:inputsy}
\eeq
Fitting the line eq.~(\ref{eq:lineformeasurement}) on the results obtained from the energy spectra, we obtain the best-fit lines shown in Figure~\ref{fig:MeasureData}, which correspond to
\beq
\texttt{s}=(4.8\pm 0.3)\times 10^{-4}\gev^{-1},\quad \texttt{y}=597\pm 5 \gev \,,\label{eq:massivesy}
\eeq
for the massive template and
\beq
\texttt{s}=(5.2\pm 0.3)\times 10^{-4}\gev^{-1},\quad \texttt{y}=606\pm 6 \gev \,,\label{eq:masslesssy}
\eeq
for the massless template.
Not surprisingly, the  values extracted using the massive template are closer to the input values than those from the massless template, although even the massless template was able to get a rough estimate of the rest frame energy values of the $bb$ system for the chosen values of $\bar{m}_{bb}$. In Figure~\ref{fig:contour}, we also show the 68\% and 95\% confidence level contours  obtained from the $\chi^{2}$ variation of the fit of the slope and intercept parameters; the result with the massive template are in the left panel and that with the massless template in the right panel. One can clearly see that the distance between the theory values and the best-fit  values for the case of the massive template (left panel) is smaller than that for the case of the massless template (right panel).

As mentioned before, \texttt{s} and \texttt{y} can be easily converted into the masses of gluino and neutralino. Based on eqs.~(\ref{eq:massivesy}) and~(\ref{eq:masslesssy}), we obtain the following measurements of the two masses:
\bea
\hbox{Massive template}:&& m_{\tilde{g}}=1042\pm 65 \hbox{ GeV},\;\;m_{\chi}^2=-159000\pm 59000 \hbox{ GeV}^2\,,\label{eq:massiveresults} \\
%-(398.7)^{2} \pm (242)^{2}
\hbox{Massless template}:&& m_{\tilde{g}}=964\pm 56 \hbox{ GeV},\;\;m_{\chi}^2=-240000\pm 41000 \hbox{ GeV}^2 \,. \label{eq:masslessresults}
%-(490)^{2}\pm (202)^{2}
\eea
We remark that the gluino mass, while quite precisely determined, is underestimated by about 20\%, with the value from the massive template being closer to the true value than the value from the massless template. The neutralino mass is poorly determined using both the massive and the massless template. Possible causes of this poor estimation will be discussed in the next subsection.

\subsection{Study of systematic effects \label{sec:improvement}}

The results in the previous subsection on the measurement of the gluino and neutralino masses are fairly good, considering the challenging circumstances of the mass measurement, particularly the fully indistinguishable character of final state particles in our chosen signal process. Despite an adequate result for the gluino mass measurement, the neutralino mass measurement is very poor; the only conclusion that one is able to draw is that the neutralino in our example process is consistent with being massless. 

\begin{table}[t]
\centering
\begin{tabular}{|cc|c|c|c|}
\cline{1-5}
 & &\multicolumn{3}{c|}{slope} \\ \cline{3-5}
 & & steeper & consistent & shallow \\ \cline{1-5}
\multicolumn{1}{|c}{\multirow{6}{*}{$y$-intercept}} &
\multicolumn{1}{|c|}{\multirow{2}{*}{larger}} & $m_{\tilde{g},\textnormal{ext}}<m_{\tilde{g},\textnormal{in}}$ & $m_{\tilde{g},\textnormal{ext}} \approx m_{\tilde{g},\textnormal{in}}$ & $m_{\tilde{g},\textnormal{ext}}>m_{\tilde{g},\textnormal{in}}$ \\
\multicolumn{1}{|c}{} & 
\multicolumn{1}{|c|}{} & $m_{\tilde{\chi}_1^0,\textnormal{ext}}^2\ll m_{\tilde{\chi}_1^0,\textnormal{in}}^2$ & $m_{\tilde{\chi}_1^0,\textnormal{ext}}^2< m_{\tilde{\chi}_1^0,\textnormal{in}}^2$ & $m_{\tilde{\chi}_1^0,\textnormal{ext}}^2\approx m_{\tilde{\chi}_1^0,\textnormal{in}}^2$ \\ \cline{2-5}
\multicolumn{1}{|c}{} &
\multicolumn{1}{|c|}{\multirow{2}{*}{consistent}} & \cellcolor{orange} $m_{\tilde{g},\textnormal{ext}}<m_{\tilde{g},\textnormal{in}}$ & $m_{\tilde{g},\textnormal{ext}} \approx m_{\tilde{g},\textnormal{in}}$ & $m_{\tilde{g},\textnormal{ext}}>m_{\tilde{g},\textnormal{in}}$ \\
\multicolumn{1}{|c}{} & 
\multicolumn{1}{|c|}{} & \cellcolor{orange} $m_{\tilde{\chi}_1^0,\textnormal{ext}}^2< m_{\tilde{\chi}_1^0,\textnormal{in}}^2$ & $m_{\tilde{\chi}_1^0,\textnormal{ext}}^2\approx m_{\tilde{\chi}_1^0,\textnormal{in}}^2$ & $m_{\tilde{\chi}_1^0,\textnormal{ext}}^2> m_{\tilde{\chi}_1^0,\textnormal{in}}^2$ \\ \cline{2-5}
\multicolumn{1}{|c}{} &
\multicolumn{1}{|c|}{\multirow{2}{*}{smaller}} & $m_{\tilde{g},\textnormal{ext}}<m_{\tilde{g},\textnormal{in}}$ & $m_{\tilde{g},\textnormal{ext}} \approx m_{\tilde{g},\textnormal{in}}$ & $m_{\tilde{g},\textnormal{ext}}>m_{\tilde{g},\textnormal{in}}$ \\
\multicolumn{1}{|c}{} & 
\multicolumn{1}{|c|}{} & $m_{\tilde{\chi}_1^0,\textnormal{ext}}^2\approx m_{\tilde{\chi}_1^0,\textnormal{in}}^2$ & $m_{\tilde{\chi}_1^0,\textnormal{ext}}^2> m_{\tilde{\chi}_1^0,\textnormal{in}}^2$ & $m_{\tilde{\chi}_1^0,\textnormal{ext}}^2\gg m_{\tilde{\chi}_1^0,\textnormal{in}}^2$ \\ \cline{1-5}
\end{tabular}
\caption{\label{tab:casedivision} Comparisons of extracted mass parameters with corresponding input values for the nine possible combinations of over-, under- or consistent estimation of the slope and the intercept of the straight line eq.~(\ref{eq:lineformeasurement}) fitted on the $(m_{bb},E_{bb}^{*})$ data. The orange table cell corresponds to the result of the fit of the data in Sec.~\ref{sec:baselinemethod}.}
\end{table}

As noted already, the measurement of $E_{bb}^{*}$ for each $\bar{m}_{bb}$ is statistically compatible with the theory value, but still results in a mass measurement that is systematically overestimated. From the fit of the data in Figure~\ref{fig:MeasureData}, the mismeasurement of $E_{bb}^{*}$ primarily implies a slope larger than that predicted by theory, which consequently implies that the extracted gluino mass is biased towards values smaller than the true mass. This bias is not particularly worrisome {\it per se}, as it is about 10\%. However, given the relation between the masses and the observables in eq.~(\ref{eq:observablestomasses}), it turns out that this underestimation of the gluino mass severely affects the neutralino mass determination.
More generally, there are nine possible cases based on under-, over-, or consistent estimations of the slope and intercept of the straight line in eq.~(\ref{eq:lineformeasurement}). The implication of each case in terms of the extracted mass parameters is summarized in Table~\ref{tab:casedivision}.

It is interesting to examine possible causes of this bias in the best-fit line of Figure~\ref{fig:MeasureData}, which also serves as a basis for possible improvements of our method. 
In order to clarify the origin of the incorrect estimation of $E^{*}$, we study the following potential sources of inaccuracy in our fits of the energy spectra:
\begin{itemize}
\item[i)] an imperfect fit of the data with the massive template eq.~(\ref{eq:massiveTemplate});
\item[ii)]  contamination due to the background;
\item[iii)] biases introduced by the event mixing subtraction;
\item[iv)] finite size of the $m_{bb}$ range used to discretize the multi-body phase-space; 
\item[v)] biases due to events selection.
\end{itemize}

For the first potential source, we recall the discussion in Section~\ref{sec:massivetemplate} where the massive template function was introduced; this template had a maximum at $E_{bb}^{*}$ only when $w\to\infty$, which corresponds to producing the gluino(s) at rest. For practical cases, $w$ is finite and the maximum of the function appears at a somewhat larger value than $E^{*}_{bb}$. On the other hand, physical energy
distributions can have the maximum at $E_{bb}^*$; in particular, for cases where $m_{bb}$ can be treated as effectively massless. Therefore, the relevant fit could result in a value that does not match the corresponding expectation. Fortunately for the case at hand, we find that $w$ is large enough to cause only a negligible shift in the peak position, i.e., such a potential mismatch is very tiny. Consequently, we do not ascribe the systematic overestimate $E_{bb}^{*}$ to the inaccuracy of the template function eq.~(\ref{eq:massiveTemplate}).

In order to see the effect of the other four potential sources of bias on the final result, we conduct a dedicated analysis for each. In each analysis, we repeat the same procedure as described in the previous section, that is to say we extract the values of $E_{bb}^{*}$ from an event sample that incorporates the effect under study. The event samples for the study of these possible effects are denoted as ``Check Sample'' (CS). We then compare the results obtained from those Check Samples to those obtained from the Original Sample (OS). The attributes of the check samples that we have considered are summarized in Table~\ref{tab:controlsamples} and are also described in the following.

\begin{table}[t]
\centering
\begin{tabular}{c|c c c c c}
 & $\Delta m_{bb}$ & Event mixing & Background & Background fit & Cuts\\
\hline
OS & 50 GeV & Yes & Included &No & Yes \\
CS I & 50 GeV & Yes & Included &Yes & Yes\\
CS II & 50 GeV & No & Included &No & Yes\\
CS III & 50 GeV & No & Not Included & - & Yes\\
CS IV & 2 GeV & No & Not Included & - & Yes \\
CS V & 2 GeV & No & Not Included & - & No \end{tabular}
\caption{\label{tab:controlsamples} Description of the original sample (OS) and several selected check samples (CS).  The width of the ranges of $m_{bb}$ for the discretization of the multi-body phase space is reported in the first column. Samples marked as event mixed are those in which the mixed event subtraction has been carried out. In those marked as ``no'', the correct pairs are identified in the event record and so eliminate the effect of combinatorial backgrounds. Samples where the background has been completely neglected are marked in the third column. For the samples where the background has been added, we report in the fourth column if we have added a template to fit the background events to the overall fit of the data . Finally, in the fifth and final column we report if selection cuts eqs.~(\ref{eq:cutforb}) through~(\ref{eq:delphicut}) have been applied to the events or not.}
\end{table}  

\begin{figure}[t]
\centering
\includegraphics[width=7.5cm]{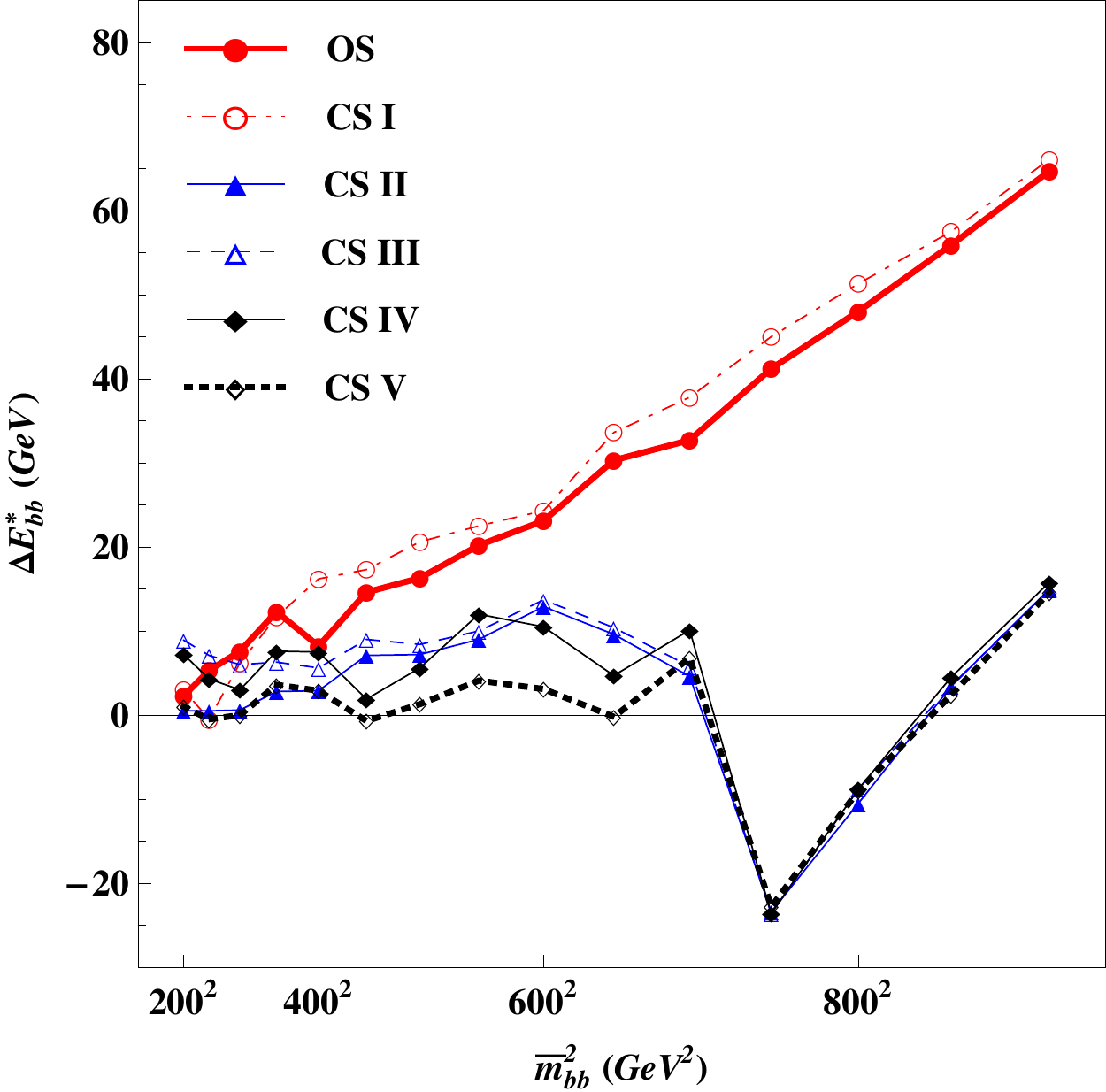}
\includegraphics[width=7.5cm]{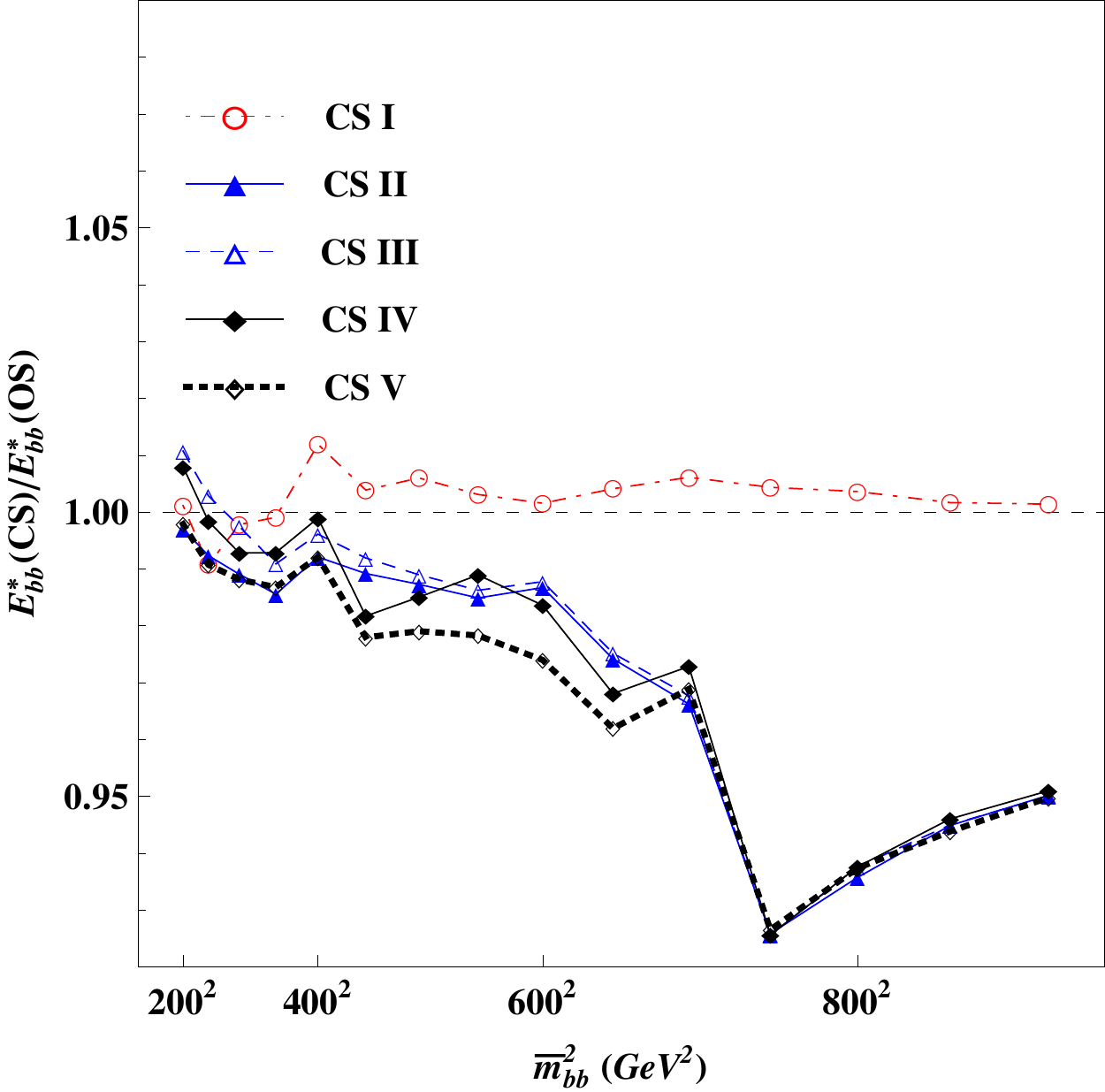}
\caption{\label{fig:controls} Comparisons of fit results from the five samples used to assess the effects of several potential sources of bias in the gluino mass determination as described in Table~\ref{tab:controlsamples}.}
\end{figure}

The first check sample enables us to find the effect of the background on the extraction of $E_{bb}^{*}$.
 We study first the pure background energy distributions in order to calibrate the template function describing them in the fit. This calibration is done for each of the $\bar{m}_{bb}$ slices. Although there are two types of backgrounds, the dominant SM background and the interference from the event mixing, we employ a single template in eq.~(\ref{eq:massiveTemplate}) to describe both of them collectively. We then repeat the fit of the energy spectra for each $\bar{m}_{bb}$,  including the template function for the backgrounds as well. 
The results in the determination of $E^{*}_{bb}$ for each $\bar{m}_{bb}$ are labeled as ``CS I'' an plotted as red open circles in Figure~\ref{fig:controls}. In this figure, the left panel shows the absolute shift of the measured $E_{bb}^*$ from the corresponding theory value for each sample. The right panel shows the ratio of the values of $E_{bb}^*$ from the check samples and the corresponding value in the original sample. We observe that the effect of background modeling is negligible for all $\bar{m}_{bb}$, and from the right panel of Figure~\ref{fig:controls}, we can in fact see that this engenders less than 1\% of the shift in $E_{bb}^*$. 

The next potential source of bias that we study is the event mixing, which is studied by the second and third check samples, denoted by CS~II and CS~III in the following.  In these samples we use the event record to identify the correct pairs of jets coming from the same gluino, and therefore we obtain the correct energy spectra without applying the mixed event subtraction.
 The two samples CS~II and CS~III differ by the inclusion of the SM background.  The results for the determination of $E_{bb}^{*}$ are reported in Figure~\ref{fig:controls} by blue filled triangles and blue open triangles, respectively. From the figure we see that the determination of $E_{bb}^{*}$ is significantly improved. In fact, in the left panel of Figure~\ref{fig:controls} we can see that a mild positive shift of the determined $E_{bb}^{*}$ values still exists, but is greatly reduced compared to what we had with the original sample.
We remark further that only minor differences are found between the results obtained from the check samples CS~II and CS~III, which can be taken as another way of confirming that the effect from background events is negligible. 
Therefore, we conclude that the effect of event mixing is a major cause of the shift that we observe in the gluino mass determination.

Next, we study the effect of the discretization of the $m_{bb}$ spectrum by taking smaller ranges for the $\bar{m}_{bb}$ window. This is performed on the check sample denoted as CS~IV. This narrow $m_{bb}$ range analysis is intended to provide an improvement to the previous two analyses where the combinatorial issues were artificially resolved using the information in the event record. In fact, the check sample CS~IV for this analysis is similar to CS~III except for the $\Delta m_{bb}$.  The results of this analysis are reported in Figure~\ref{fig:controls} by black filled rhombuses, which suggests that the effect of discretization of the $m_{bb}$ spectrum is negligible. This is not surprising in light of the following observation:  one can easily figure out that, for a given nominal value of $\bar{m}_{bb}$ the the $E_{bb}^*$ from eq.~(\ref{eq:lineformeasurement}) for $(\bar{m}_{bb}+25)\gev$ $(\bar{m}_{bb}-25)\gev$ is at most about 15~GeV larger (smaller) than the $E_{bb}^*$ for the nominal $\bar{m}_{bb}$. The absolute size of this shift is already quite small and is further reduced by the fact that for each $m_{bb}$ range we observe the sum of all the contributions below and above $\bar{m}_{bb}$. In all, we expect a very small net effect, making the discretization of the $m_{bb}$ spectrum in increments of 50~GeV suitable for the precision sought. 

Finally, we study the bias induced by the selection cuts. To assess their effect, we produce a sample along the line of CS~IV, but being fully inclusive in the signal phase-space. The result of fits performed on the energy spectra from this sample are reported in Figure~\ref{fig:controls} by black open rhombuses. The use of a fully inclusive sample gives $E_{bb}^*$ from the fits that agree with the theory predictions within few percents up to $\bar{m}_{bb}=650$ GeV. 

For the check samples II, III, IV and V we remark that the agreement of the fit results with the theory value deteriorates as one gets closer to the endpoint of the range covered by $m_{bb}$. We observe that good agreement is retained up to $\bar{m}_{bb}=650$ GeV, which is in the falling tail of the $m_{bb}$ distribution, as apparent from Figure~\ref{fig:puresig}. 
We suspect that the mismatch of the fitted $E_{bb}^{*}$ and the theory values is connected to the massive template becoming less accurate in fitting to the data. Indeed, in Figure~\ref{fig:MeasureData} we can see that the error estimation on the fitted $E_{bb}^{*}$ becomes larger for $\bar{m}_{bb}\geq 650$ GeV. In a realistic application of our mass measurement method, we would not know up to what precise value of $m_{bb}$ the massive template can be trusted. However, it is clear that the values of $m_{bb}$ for which the extracted $E_{bb}^{*}$ from the fit comes with a large error should be avoided. We remark that in Figure~\ref{fig:MeasureData}, all the fit results for $m_{bb}\geq 650\gev$ have a significantly large error, thus clearly signaling a transition to a region of $m_{bb}$ where the fit template eq.~(\ref{eq:massiveTemplate}) can no longer be trusted. A more detailed investigation of this transition boundary is beyond the scope of our paper and we instead refer to Ref.~\cite{Agashe:2015ike} for a more systematic study of it.

\subsection{Improving the mass measurement using the  $m_{bb}$ endpoint}
\begin{figure}[t]
\centering
\includegraphics[width=7.5cm]{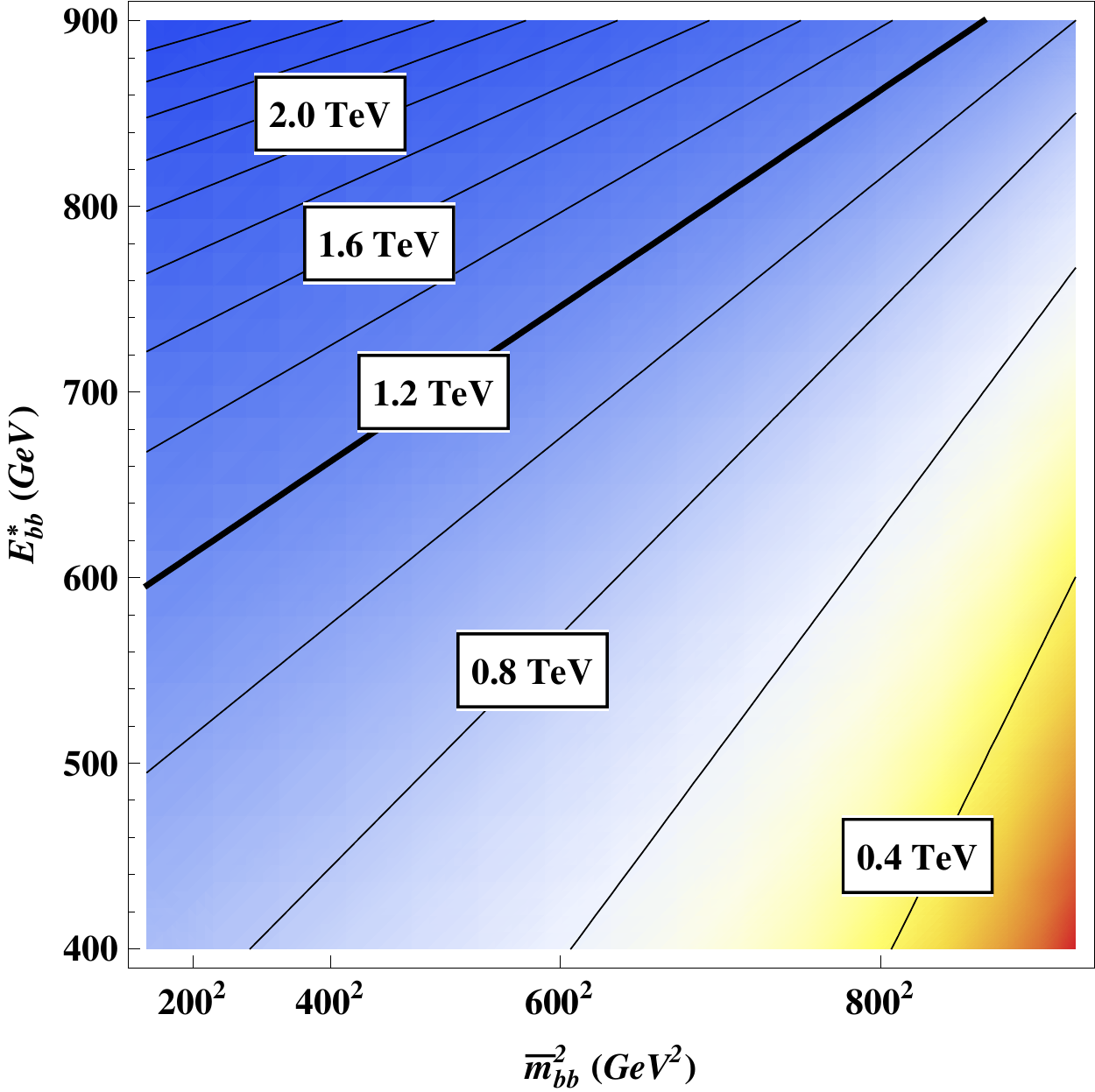}
\caption{\label{fig:contoursConstFit} The functional dependence of eq.~(\ref{eq:modfitter}) for various gluino masses. The $m_{bb}$ endpoint is set to be 1.1 TeV. The black solid line denotes the case where $m_{\tilde{g}}$ is identical to that of our study point. }
\end{figure}
 In this section, we discuss a possible improvement of the mass measurement with the aid of the kinematic endpoint of the dijet invariant mass distribution. The point is that without prior knowledge of the masses, it is not possible to say whether the measurement eq.~(\ref{eq:massiveresults}) has a bias. However, we can devise a check and an improvement of the obtained measurement using an independent observable. To this end, we study a possible combination of the result of fits to energy spectra with the measurement of the endpoint of the invariant mass distribution of the pairs of $b$-jet, which we denote as $m_{bb}^{\max}$. This observable is a simple function of the gluino and neutralino masses:
\bea
m_{bb}^{\max}=m_{\tilde{g}}-m_{\chi}\,,\label{eq:mbbmax}
\eea
and is expected to be very useful in combination with the results of fits to energy spectra. In fact,
 eq.~(\ref{eq:lineformeasurement}), in light of eq.~(\ref{eq:mbbmax}), can be rewritten as 
\bea
E_{bb}^*= m_{bb}^{\max}- \frac{\left(m_{bb}^{\max}\right)^{2}-\bar{m}_{bb}^{2}}{2m_{\tilde{g}}}\,,\label{eq:modfitter}
\eea 
which is a straight line in the plane $(E_{bb}^{*},\bar{m}_{bb}^{2})$ described by just one free parameter. This should be compared to the previous equation we used to find the masses, eq.~(\ref{eq:lineformeasurement}), where there are two independent parameters: the slope and the constant term of the straight line.

\begin{figure}[t]
\centering
\includegraphics[width=7.1cm]{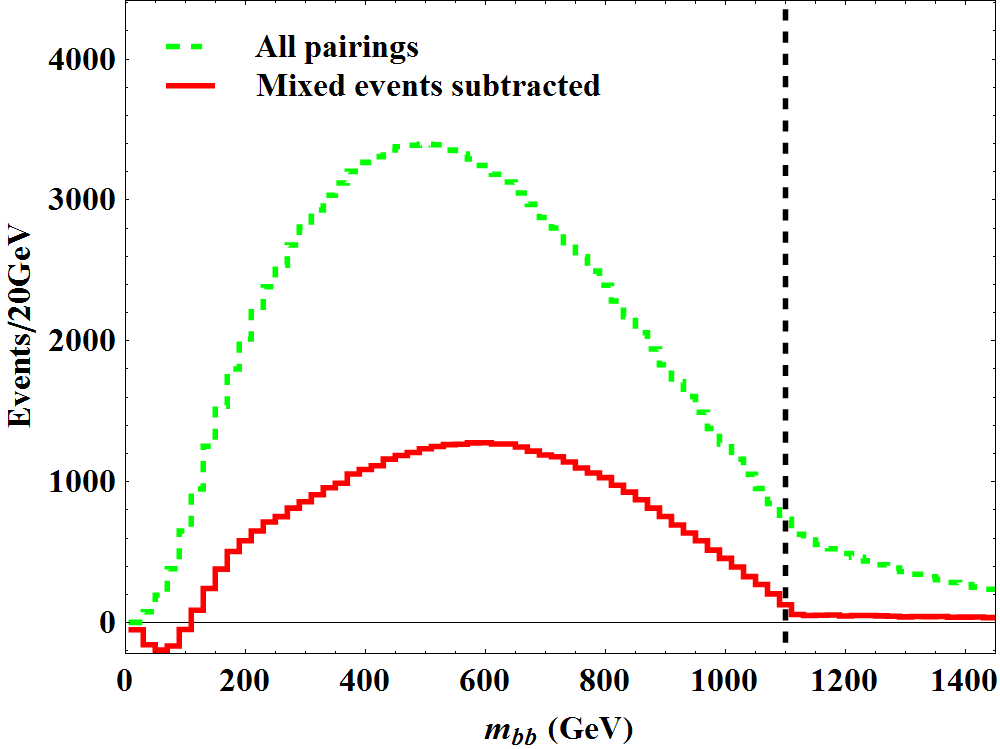} \hspace{0.5cm}
\includegraphics[width=7cm]{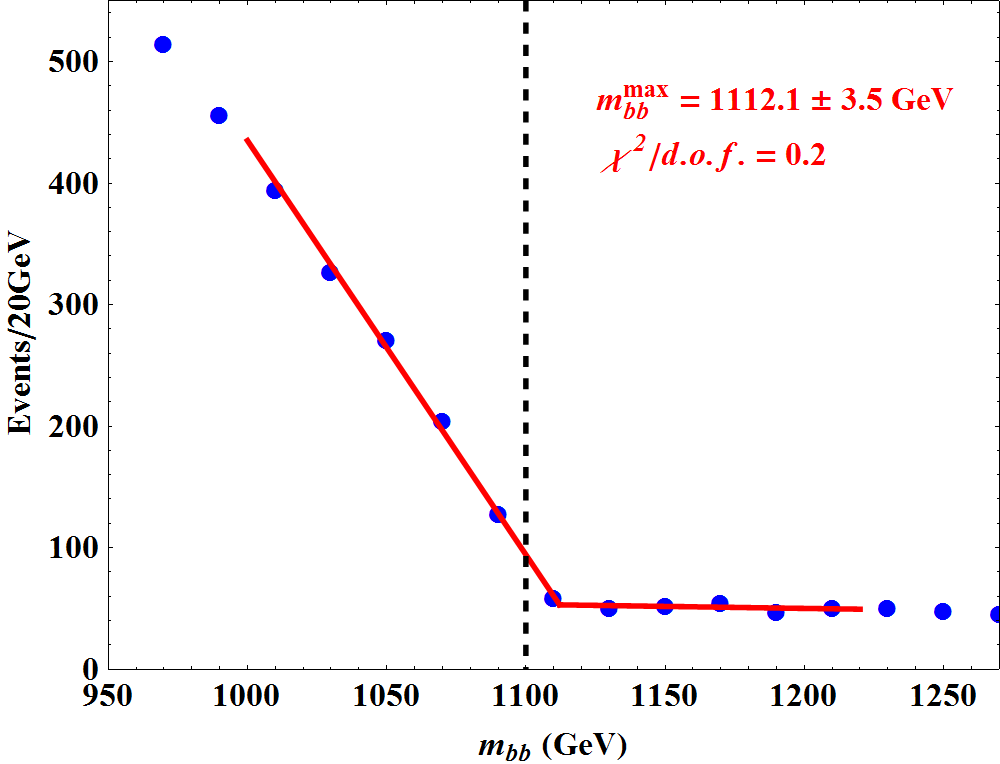}
\caption{\label{fig:endfit} The left panel shows the di-$b$-jet invariant mass distributions that are normalized to an integrated luminosity of 3 ab$^{-1}$. 
Both signal and background events are plotted with selection cuts in Section~\ref{sec:eventselection} imposed.
The right panel shows the endpoint extraction by fitting the $m_{bb}$ data between 1000 and 1200 GeV with the template in eq.~\eqref{eq:mbbfitter}. }
\end{figure}

If the $m_{bb}$ endpoint is assumed to be well-measured, we can use eq.~(\ref{eq:modfitter}) to more accurately fit our results to a line in the plane $(m_{bb}^2, E_{bb}^*)$. This relation is shown in Figure~\ref{fig:contoursConstFit} for $m_{bb}^{\max}=1.1$~TeV.
The determination of kinematic endpoints is a well-established technique, as experimental collaborations have adopted the idea for precision measurement (e.g. Ref.~\cite{Chatrchyan:2013boa}).
Several phenomenological works have developed techniques of detecting kinematic endpoints or edges in one-dimensional data~\cite{Agashe:2010tu,Curtin:2011ng} and even in multi-dimensional data~\cite{Debnath:2015wra}.
To construct the template for the endpoint extraction, we  closely look at the dijet invariant mass distribution. 
As explicitly shown in Figure~\ref{fig:puresig}, the combinatorial background is a major challenge, and we again conduct the mixed event subtraction with a signal-background-combined event sample.
The left panel of Figure~\ref{fig:endfit} exhibits the distribution constructed by all possible pairs (green dashed histogram) and the distribution after the mixed event subtraction (red solid histogram). 
To guide the location of the theoretic endpoint, we lay a black, vertical dashed line. 
Since the full shape of the $m_{bb}$ distribution in this three-body decay is well-known (see, for example, Refs.~\cite{BK,Cho:2012er}), one could employ the relevant analytic expression to fit the signal portion. 
However, considering potential shape distortion by the mixed event subtraction, non-trivial spin effects, matrix element etc. (although not significant), we instead focus on local information in the vicinity of the kinematic endpoint and describe it by a straight line. 
We also observe that the residual distribution beyond the endpoint can be also accommodated by a single straight line. 
We therefore introduce the following template $g(m_{bb})$ to extract the maximum $m_{bb}^{\max}$:
\bea
g(m_{bb})=s_1(m_{bb}-m_{bb}^{\max})+s_2m_{bb}+c\,,\label{eq:mbbfitter}
\eea
where $s_{1,2}$, $c$, and $m_{bb}^{\max}$ are fit parameters responsible for slopes, $y$-intercept, and the endpoint, respectively. 
We remark that similar strategies have been implemented by experimentalists (e.g., Ref.~\cite{Chatrchyan:2013boa}) and in many phenomenological studies (e.g., Ref.~\cite{Cho:2007qv}).

Our fit result with the above template is reported in the right panel of Figure~\ref{fig:endfit}. 
The data points (blue dots) in-between 1000 and 1200 GeV are taken for the fit, and the best-fit lines are represented by two red straight lines.
The extracted endpoint value is 
\bea
m_{bb}^{\max} = 1112.1 \pm 3.5 \hbox{ GeV},
\eea
where the quoted error is again a statistical uncertainty.
The result is quite precise as expected. 
Using this extracted $m_{bb}^{\max}$ and the relation in eq.~\eqref{eq:modfitter} as a template for the fit of the $(m_{bb}^2, E_{bb}^*)$ data points in Figure~\ref{fig:MeasureData} along with eq.~(\ref{eq:mbbmax}), we obtain a mass measurements:
\bea
m_{\tilde{g}}=1231 \pm 30 \hbox{ GeV}, \;\;\;m_{\chi}= 119 \pm 30 \hbox{ GeV}\,.\label{eq:massmeasurementmbb}
\eea
In this case, as well as for the analysis in the previous sections, the fit is performed between $\bar{m}_{bb} = 200$ and $\bar{m}_{bb} = 650$ GeV, with the data points obtained from the massive template. 
This result is more accurate and in better agreement with the expected values than what we obtained in eq.~(\ref{eq:massiveresults}) using only energy distributions. Therefore, we conclude that, depending on the accuracy with which the $m_{bb}$ endpoint can be experimentally determined, the addition of information from the $m_{bb}$ endpoint can grant a very significant improvement to our results obtained from only the energy spectra.

\section{Summary and Conclusions\label{conclude}}
In this paper, we have discussed how to use the energy spectra of visible decay products for the measurement of masses of ``parent"
particles in semi-invisible {\em multi} (i.e., more than 2)-body decays. The results are an extension of previous results regarding the properties of the energy distribution of the massless decay products in a two-body decay. In particular, we extended the results for two-body decays by discretizing the multi-body phase-space and considering it as the combination of multiple two-body decays, as suggested by the recursive factorization formula for the multi-body phase space.

The fictitious two-body systems that are involved in the recursive factorization of phase-space are necessarily massive. Therefore, we utilized the results of a companion paper \cite{Agashe:2015ike} on the description of energy spectra of massive decay products in two-body decays. Armed with these results, we 
should be
able 
to fit the energy spectra of the visible part of the fictitious two-body decay and extract an estimate of the masses involved in the process using the results of these fits.

A particularly challenging aspect of our analysis had to do with the fact that the decaying particle (the particle of interest) is typically produced in association with other particles. In this situation, it is possible that some of the ``child" particles from the decay of the parent are identical to
those contained in the {\em rest} of the event, which means that in the process of reducing the multi-body final state of the parent decay to one with fewer bodies, the particle pairings that we perform may unintentionally include particles which have nothing to do with the mass measurement at hand.
In particular, the parent particles are often produced {\em in pairs}: if the two parents in each event undergo the same decay process,
then it is clear that this combinatorial background is inevitable.
These particles extraneous to the decay potentially hamper the mass measurement of the parent,
and thus the contamination they add must be addressed.
Most of the general discussion above can be succinctly illustrated by the consideration of a suitable example.
Furthermore, tackling a concrete example enables us to quantify the quality of the mass measurement that one can achieve using our method.
With these goals in mind, 
we studied in detail the production of a pair of gluinos in a supersymmetric model, where R-parity is conserved, and in which the gluinos directly decay to $b \bar{b} \tilde{\chi}_1^0$ (a three-body decay) via an {\em off}-shell bottom squark.
To this end, we simulated events for the process
\bea
pp\to \tilde{g}\tilde{g} \to bbbb+\misse \nonumber\,,
\eea
including the relevant SM contribution.
We identified selection cuts to remove the SM 
backgrounds to a level that further clears a path toward the successful determination of the masses of the new physics states. Simultaneously, we attempted to minimize the changes in the shape of the energy spectra caused by these event selection criteria.

Our actual analysis then starts with forming {\em all} possible pairs of bottom quarks in each event that passed the selection in eqs.~(\ref{eq:cutforb})-(\ref{eq:delphicut}), from which we then obtained a distribution of the energy of the $b$ quark pairs, $E_{bb}$. This distribution includes the contribution from pairs formed by bottom quarks not originating from the same decay, i.e., the {\em wrong}
combinations. 
Note that the final state of each decay in our chosen process is made of two (visible) indistinguishable particles, and so the pollution from the combination of particles not coming from the same decay is even more severe; in particular,
each event gives 6 combinations of two bottom quarks, out of which only 2 are correct, cf. the case of distinct
particles $a$ and $b$ from each decay, which would give  2 correct ones out of 4 combinations.
To remove this adulteration of the event sample, we subtracted an estimate that we obtained using the event mixing technique described in Sec.~\ref{sec:eventmixing}. This estimate was obtained from pairs of $b$ quarks taken from {\it different} events, and we showed that the contribution from pairs of $b$ quarks not coming from the same gluino can thus be effectively removed, as seen in Figure~\ref{fig:energyMESS}. 
In other words, this method has a natural tendency to maintain the shapes of the distributions that we need to analyze to carry out our measurement.

The energy spectra, once effectively rid of the contribution from pairs of $b$ quarks coming from different gluinos, were fitted in a region around their peak with the function eq.~(\ref{eq:massiveTemplate}), which is taken from our work Ref.~\cite{Agashe:2015ike} and briefly described in Sec.~\ref{sec:massivetemplate}.
For comparison, these energy spectra were also fitted with the function eq.~(\ref{eq:masslessTemplate}), which was used in our original paper on the peak of energy spectra of mass{\em less} particles. The comparison of the results from fits with the two functions highlights the improvement achieved by our new result for massive decay products. The better description of the energy spectra with the function eq.~(\ref{eq:massiveTemplate}) can be seen in the comparison of the $\chi^{2}$ for the various fits performed, as reported in Figure~\ref{fig:chiratio}.

The result of the fits of the energy spectra is the extraction of the function parameter $E_{bb}^{*}$, which is exactly the energy of the system of the two $b$ quarks in the rest frame of the gluino that generated the two $b$ quarks. The determination of $E_{bb}^{*}$ is the core of our procedure as this value is connected to the masses of the gluino and the neutralino via eq.~(\ref{eq:lineformeasurement}):
\bea
E_{bb}^{*}= \frac{m_{\tilde{g}}^{2}-m_{\chi}^{2}+\bar{m}_{bb}^{2}}{2m_{\tilde{g}}} \nonumber
\eea
for a pair of $b$ quarks of mass $\bar{m}_{bb}$.
The results of the extraction of $E_{bb}^{*}$ for several choices of the mass of the two $b$ quark system were shown in Figure~\ref{fig:MeasureData}. The determination of $E_{bb}^{*}$ for each $m_{bb}$ was fitted using the straight line  eq.~(\ref{eq:lineformeasurement}) given above. This fit is essentially our mass measurement, as the gluino mass corresponds to the slope of the line and the neutralino mass to the constant term of the straight line. The resulting mass measurement was given in eq.~(\ref{eq:massiveresults}), which was found to be within 20\% of the gluino mass and a rather poor determination of the neutralino mass.
This inaccuracy of the mass measurement is mainly due to the fact that in each fit of energy spectra, we tend to overestimate the energy $E_{bb}^{*}$. In Sec.~\ref{sec:improvement}, we studied several possible causes for this error
and (in the end) identified the modest shape changes due to the mixed event subtraction as the primary source. 

In order to improve the mass measurement, we then studied how including information about the endpoint location in the $m_{bb}$ distribution altered the quality of the mass measurements. If well-measured, this quantity should correspond to the mass difference between the gluino and the neutralino, and hence is expected to aid in the measurement of these two masses. 
Indeed,
a much more accurate measurement was obtained using information from this observable, as shown in  eq.~(\ref{eq:massmeasurementmbb}).

As the analysis we have discussed in this work used parton-level events, we finally make a brief comment on the potential impact of the effects that would be necessary to include in a fully realistic study beyond this 
%
%parton-
%
level ($e.g.$ parton showering, hadronization, detector response, etc.).
Since our target final state
%
%object 
%
is the bottom quark jet, present experience with such objects 
in the LHC experiments make us confident that our conclusions would not change significantly if we went beyond a parton-level analysis.
In fact, we can look at similar analyses with $b$-jet energy spectra and notice that detector effects and the improvement of event generation beyond the parton-level induce small effects that are well under control for the level of accuracy of the measurements discussed in this work. 
For example, Ref.~\cite{Agashe:2012bn} performed a detector-level study on the $b$-jet energy spectrum from the top quark decay to measure the top quark mass, and found that the extracted $E^*$ is in good agreement with the corresponding input value within the associated uncertainty. 
More dedicated study on the higher order effects in the $b$-jet energy spectrum from top quark decay has been conducted in Ref.~\cite{Agashe:2016bok} and demonstrate that these effects are negligible for our purpose here.
%
%the accuracy of the measurements discussed in this work.  
%
Finally, the CMS collaboration adopted the proposal in Ref.~\cite{Agashe:2012bn} and measured the top quark mass using the $b$-jet energy spectrum~\cite{CMS:2015jwa}. The measured top quark mass from the extracted $E^*$ was quite consistent with the world-average value; in particular,  
%
%within the associated uncertainty and 
%
despite
the uncertainties associated with $b$-jet energy, a
%
%(roughly) 
%
percent-level precision 
%
%uncertainty on
%
measurement of 
the top quark mass
was obtained. 
From all these considerations, we anticipate that for the signal discussed in this work the reconstructed $b$-jets do not 
develop 
%
%incur 
%
a substantial systematic bias in their energy distribution, and the relevant 
%
%systematic 
%
uncertainties can be under control. 
%We leave more detailed investigation for future work. 

In conclusion, we have demonstrated the use of energy distributions, and in particular, the region close to their peaks, to measure the masses of new physics particles involved in a single-step multi-body decay. Taking the example of gluino production and decay in a R-parity conserving supersymmetric model, we have found that using only visible decay products of the gluino decay $\tilde{g}\to bb\chi$, it is possible to measure with good accuracy the masses of both the gluino and neutralino, with the best results being obtained when {\em both} the energy and the invariant mass distributions of the pairs of $b$ quarks are used. 
Rather strikingly,
the mass measurement technique that we discussed does not actually use information about the missing momentum, except only for event selection purposes. 
The example that we considered also required a proper removal of the effect from pairs of $b$ quarks not coming form the same gluino. To this end, we have shown that the event mixing technique is especially well-suited. 
We anticipate that our general methodology here can be applied to mass measurements in other processes.

%%%%%%%%%%%%%%%%%%%%%%%%%%%%%%%%%%%%%%%%%%%%%%%%%%%%%%%%%%%%%%%%%%%%%%%%%%%%%%%%%%%%%%%%%%%%%%%%%%%%%%%%%%%%%%%%%%%%%%%%%%%%%%%%%%%%%%%%%%%
%%%%%%%%%%%%%%%%%%%%%%%%%%%%%%%%%%%%%%%%%%%%%%%%%%%%%%%%%%%%%%%%%%%%%%%%%%%%%%%%%%%%%%%%%%%%%%%%%%%%%%%%%%%%%%%%%%%%%%%%%%%%%%%%%%%%%%%%%%%

\section*{Acknowledgements}
We would like to thank Michele Papucci for suggesting the idea of phase-space slicing, Konstantin Matchev and Myeonghun Park for suggesting and discussing the idea of the mixed event subtraction scheme, and Won Sang Cho for useful discussions.
This work was supported in part by NSF Grant No.~PHY-0968854 and by the Maryland
Center for Fundamental Physics.
%
%-0652363. 
%
The work of K.~A., R.~F., and K.~P.~W. was supported in part by NSF Grant No. PHY-1315155. 
The work of R.~F. is also supported by the NSF Grant No.~PHY-0910467. %
The work of D.~K. is supported in part by U.S. Department of Energy Grants DE-FG02-97ER41029.
D.~K. also acknowledges the support from
the LHC Theory Initiative graduate and postdoctoral fellowship (NSF Grant No.~PHY-0969510).

\appendix

\section{\label{sec:AppendixA} Event mixing technique with a signal plus background sample}
We begin with the total number of pairings of $b$-jets (denoted by $N_{NC}$) and with the data sample consisting of $n_s$ signal events and $n_b$ background events. Remembering the fact that there are six possible pairings out of four $b$-jets in each event, we have
\bea
N_{NC} = 2n_s + 2n_b + 4n_s + 4n_b
\eea
where the last two terms represent the total number of wrong pairings of $b$-jets denoted by $$N_{WC}=4(n_s+n_b)\,.$$

As mentioned before, our unawareness of which are signal and which are background events precludes us from having the event mixing (symbolized by $\otimes$) only between signal events or background events. Therefore, if we perform the event mixing procedure on the total $(n_s+n_b)$ events, we then have signal-signal mixing, signal-background mixing, and background-background mixing. Since there are four $b$-jets in each event, a {\it single} event mixing enables us to have up to $4\time 4=16$ $b$-jet pairings. Of course, for practical purposes, one could use a subset of these 16 pairs, say $m$ pairs. Now that $m$ is given by a common prefactor for every single event mixing, we then need to know the number of event mixings. The total number of signal-signal mixing is evaluated by the two-combination in the binomial coefficients:
\bea
n_s\otimes n_s = \binom{n_s}{2}=\frac{n_s(n_s-1)}{2}, \label{eq:sxs}
\eea
and so is that of background-background mixing:
\bea
n_b\otimes n_b = \binom{n_b}{2}=\frac{n_b(n_b-1)}{2}. \label{eq:bxb}
\eea
Likewise, the number of signal-background mixing, i.e., interference, is expressed as follows:
\bea
n_s\otimes n_b = \binom{n_s}{1} \times \binom{n_b}{1} = n_s n_b. \label{eq:sxb}
\eea

Suppose that we use $m$ mixed pairings out of 16 mixed pairings in each event mixing. Denoting by $N_{MC}$ the sum of eqs.~(\ref{eq:sxs}),~(\ref{eq:bxb}), and~(\ref{eq:sxb}), we eventually use $m\cdot N_{MC}$ $b$-jet pairs to estimate the distribution from the $N_{WC}$ wrong pairs formed in the same event distribution.\footnote{Note that the only thing that we know is $n_s+n_b$, neither $n_s$ nor $n_b$ separately.} Since in general $N_{WC}\neq N_{NC}$, each mixed pair should be re-weighted in making the ``different events'' distributions, so to match the contribution of $N_{WC}$ wrong pairs. Keeping this issue of the normalization in mind, let us first see how the wrong $b$-jet pairings in the signal events  can be treated by the mixed event subtraction even in presence of a small background.  Since the total number of mixed pairings is normalized to $N_{WC}$, the presence of the background affects the weight of the ``different event'' distributions made by just signal events. In fact, in the ``different event'' distribution, once normalized so as to match the contribution of the $N_{WC}$ wrong pairs, the fraction of $b$-jet pairs stemming from the signal-signal mixing is given by  eq.~(\ref{eq:sxs}) times a rescaling factor to match the normalization to $N_{WC}$, that is
\bea
\frac{n_s(n_s-1)}{2}\times\frac{4(n_s+n_b)}{n_s(n_s-1)/2+n_s n_b+n_b(n_b-1)/2} \approx 4n_s \left(1- \frac{n_{b}}{n_s}\right) \label{eq:ssportion} 
\eea
where the common prefactor $m$ is omitted for simplicity and the approximation is done with the assumptions of $n_s \gg n_b$ and $n_s \gg 1$. The implication of this result is that the combinatorial background (caused by the signal itself) can be almost completely eliminated with the event mixing technique thanks to the dominance of the signal assumed throughout in this paper. In the total ``different event'' distribution reweighed as to match the expected number of wrong pairing $N_{WC}$ we find that the fraction of $b$-jet pairs from the background-background mixing is
\bea
\frac{n_b(n_b-1)}{2} \times \frac{4(n_s+n_b)}{n_s(n_s-1)/2+n_s n_b+n_b(n_b-1)/2} \approx 4n_b\left( \frac{n_{b}}{n_{s}}-\frac{1}{n_s} \right). \label{eq:bbportion}
\eea
In the same distribution we find that the fraction of $b$-jet pairs from the signal-background mixing is
\bea
n_sn_b \times \frac{4(n_s+n_b)}{n_s(n_s-1)/2+n_sn_b+n_b(n_b-1)/2} \approx 8n_b \left( 1-\frac{n_b}{n_s}+\frac{1}{n_s} \right). \label{eq:bsportion}
\eea

Comparing eq.~(\ref{eq:bbportion}) and eq.~(\ref{eq:bsportion}), we conclude that the effect of ``interference'' is, in general, more important than the contribution from the background-background mixing, as also argued in the main text. Besides corroborating the argument given in the main text, these equations quantify more precisely the effect of background, which becomes more important when the mass measurement strategy described in this paper is applied to situations where $S/B$ is less favorable than that in our example process.

%%%%%%%%%%
%%%%%%%%%%    References
%%%%%%%%%%

\end{document}